\documentclass[]{imsart}

\usepackage{graphicx,color}
\usepackage{caption}
\usepackage{subcaption}
\usepackage{array}
\usepackage{amssymb}
\usepackage{amsfonts}
\usepackage{multirow}
\usepackage{bigstrut}
\RequirePackage[OT1]{fontenc}
\RequirePackage{amsthm,amsmath}

\RequirePackage{natbib}
\RequirePackage[colorlinks,citecolor=blue,urlcolor=blue]{hyperref}
\usepackage{slashbox}
\startlocaldefs

\DeclareFontFamily{OT1}{pzc}{}
\DeclareFontShape{OT1}{pzc}{m}{it}{<-> s * [1.10] pzcmi7t}{}
\DeclareMathAlphabet{\mathpzc}{OT1}{pzc}{m}{it}
\newcommand{\rval}{\mathpzc{r}}

\newcommand{\vecM}{{\bf{M}}}
\newcommand{\vecm}{{\bf{m}}}

\newcommand{\vecR}{{\bf{R}}}
\newcommand{\vecr}{{\bf{r}}}

\newcommand{\vecX}{{\bf{X}}}
\newcommand{\vecY}{{\bf{Y}}}

\newcommand{\Fraud}{{\cal{F}}}

\newcommand{\vecsigma}{{\sigma}}

\newcommand{\cov}{ {\rm{cov}} }

\newcommand{\var}{ {\rm{var}} }

\newcommand{\calG}{{\cal{G}}}
\newcommand{\calS}{{\cal{S}}}

\newcommand{\vbold}{\mbox{\boldmath${v}$}}

\newcommand{\sigmagc}{$\sigma^2_{\gamma,c}$ }
\newcommand{\sigmagum}{$\sigma^2_{\gamma,1}$ }
\newcommand{\sigmagdois}{$\sigma^2_{\gamma,2}$ }
\newcommand{\sigmagtres}{$\sigma^2_{\gamma,3}$ }
\newcommand{\bgamma}{\mbox{\boldmath{$\gamma$}} }

\newcommand{\bphi}{\mbox{\boldmath{$\phi$}} }
\newcommand{\bpsi}{\mbox{\boldmath{$\psi$}} }
\newtheorem{theorem}{Theorem}
\newcommand{\I}{ {\rm{I}}}

\endlocaldefs

\begin{document}

\begin{frontmatter}

% "Title of the paper"
\title{Analysis of Aggregated Functional Data from Mixed Populations with Application to Energy Consumption}
\runtitle{Analysis of Aggregated Functional Data}

% indicate corresponding author with \corref{}
% \author{\fnms{John} \snm{Smith}\corref{}\ead[label=e1]{smith@foo.com}\thanksref{t1}}
% \thankstext{t1}{Thanks to somebody}
% \address{line 1\\ line 2\\ printead{e1}}
% \affiliation{Some University}

\begin{aug}
\author{\fnms{Amanda} \snm{Lenzi}\thanksref{t1}\ead[label=eA]{amle@dtu.dk}},
\author{\fnms{Camila} \snm{P. E. de Souza}\thanksref{t3}\ead[label=ecamila]{camila.souza@stat.ubc.ca}},
\author{\fnms{Ronaldo} \snm{Dias}\thanksref{t2}\ead[label=eR]{dias@ime.unicamp.br}},
\author{\fnms{Nancy} \snm{Garcia}\thanksref{t2}\ead[label=eNG]{nancy@ime.unicamp.br}}
\and
\author{\fnms{Nancy} \snm{E. Heckman}\thanksref{t3}
\ead[label=e2]{nancy@stat.ubc.ca}}
%\ead[label=u1,url]{http://www.foo.com}}

\thankstext{t1}{Research supported by CAPES and ELAP/Canadian Bureau for International Education}
\thankstext{t2}{Research supported by CNPq grants 302755/2010-1, 476764/2010-6 and 302182/2010-1 }
\thankstext{t3}{Research supported by the National Science and Engineering Research Council of Canada.}
\runauthor{Lenzi, A. et al.}

\affiliation{Technical University of Denmark\thanksmark{m11} , University of British Columbia\thanksmark{m2} and University of Campinas\thanksmark{m1}}

\address{Technical University of Denmark\thanksmark{t1}\\
Matematiktorvet, 303 B\\
DK-2800 Kgs. Lyngby Denmark\\
\printead{eA}\\ }
%\phantom{E-mail:\ }}
%\printead*{e2}}

\address{University of British Columbia\thanksmark{t3}\\
3181- 2207 Main Mall\\
Vancouver, BC, Canada, V6T 1Z4\\
\printead{e2}\\
\printead{ecamila}}
%\printead{u1}}

\address{University of Campinas\thanksmark{t2}\\
Rua S\'ergio Buarque de Holanda, 651\\
Cidade Universit\'aria ``Zeferino Vaz"\\
Campinas, S\~ao Paulo, Brazil, CEP 13083-859 \\
\printead{eR}\\
\printead{eNG} }
%\phantom{E-mail:\ }}
%\printead*{e2}}

\end{aug}

\begin{abstract}
Understanding the energy consumption patterns of different types of consumers is essential in any planning of energy distribution.   However, obtaining consumption information for single individuals is often either not possible or too expensive. Therefore, we consider data from aggregations of energy use, that is, from sums of individuals' energy use, where each individual falls into one of $C$ consumer classes.   Unfortunately, the exact number of individuals of each class may be unknown: consumers do not always report the appropriate class, due to various factors including differential energy rates for different consumer classes. We develop a methodology to estimate the expected energy use of each class as a function of time and the true number of consumers in each class.  We also provide some measure of uncertainty of the resulting estimates.  To accomplish this, we assume that the expected consumption is a function of time that can be well approximated by a linear combination of B-splines.  Individual consumer perturbations from this baseline are modeled as B-splines with random coefficients. We treat the reported numbers of consumers in each category as random variables with distribution depending on the true number of consumers in each class and on the probabilities of a consumer in one class reporting as another class. We obtain maximum likelihood estimates of all parameters via a maximization algorithm.  We introduce a special numerical trick for calculating the maximum likelihood estimates of the true number of consumers in each class. We apply our method to a data set and study our method via simulation.

\end{abstract}

%\begin{keyword}[class=AMS]
%\kwd[Primary ]{}
%\kwd{}
%\kwd[; secondary ]{}
%\end{keyword}

%\begin{keyword}
%\kwd{}
%\kwd{}
%\end{keyword}

\begin{keyword}
\kwd{aggregated functional data, smoothing, nonparametric regression, energy consumption}
\end{keyword}

\end{frontmatter}

\section{Introduction}

Efficient distribution of energy is a problem of vital importance to
electric companies around the world.  An important goal in energy
distribution is to not overload the energy distribution system.  To
prevent overload, distribution networks typically have been designed
to handle the maximum demand.  An alternative and more efficient
strategy to not only reduce the chance of overload but also to
maximize the use of existing equipment is to redistribute energy to
consumers so that demand is fairly constant over time.  Understanding
the typical energy use pattern of each type of consumer is essential
for this plan.

In most regions, each energy consumer belongs to one of several classes,
such as residential, commercial or industrial, and each consumer class
has a different pattern of energy consumption.  For example, in
Brazil, residential consumers appear to have a spike in electricity
consumption from 6 pm to 8 pm, due partially to the use of electric
showers after the workday.  Commercial and industrial consumers appear
to have high electricity consumption between 8 am and 6 pm.

One way to determine consumer energy use patterns is to monitor the
energy use of each consumer in a large sample composed of different
types of consumers.  However, obtaining consumer-level data is costly.
Furthermore, consumer-level data is extremely variable.  On the other
hand, aggregated data are readily available from subregions,
specifically from transformers that decrease voltage and redistribute
energy to consumers.  Samples of daily electric load of transformers
constitute our data set.  Figure \ref{fig:data.example} shows the data
collected from a transformer which redistributes energy to 41
consumers, with measurements recorded every fifteen minutes.  The plot
gives five total energy consumption curves in units of
kilo-volt-amperes of the 41 consumers, one for each weekday in the
period June 21 to June 27, 2002, inclusive.  We see that the energy
consumption curves follow a clear pattern, with little variability
from day to day.  For the data presented in Figure
\ref{fig:data.example}, we know that 7 consumers report that they are  commercial consumers
and 34 report that they are
residential consumers.  Residential consumers are of two types: those who receive monophasic electrical power and those who
receive biphasic electrical power.   Of the 34 residential consumers,  5 reported being monophasic and 29 reported being
biphasic.  However, these counts may
not be accurate -- consumers do not always report the appropriate
class, due to various factors including differential energy rates for
different consumer classes.

Our data set is from Companhia Paulista de For\c ca e Luz (CPFL) Energia, a
company that distributes electric energy in Southeast Brazil.  We
analyze data as in Figure \ref{fig:data.example}, but from three
transformers.  The three transformers serve a total of 168 consumers,
of whom 155 reported being residential customers and 13 reported being
commercial customers.    See Table \ref{tab:dataset}, which
contains reported counts and our estimates of true counts of consumer
types.

Our goal is to use this aggregated energy use data set along with the reported number of consumers in each class to estimate the true number of consumers in each class and to estimate each class's expected energy consumption as a function of time.
Our  proposed framework is to assume that each expected consumption pattern is a smooth curve that can be well approximated by a
function which is a linear combination of B-splines.  Individual consumer perturbations from this baseline are modeled as B-splines with random
coefficients.   We treat the reported numbers of consumers  in each category as random variables
with distribution depending on the true numbers of consumers in each class and on the probabilities of a consumer in one class reporting as another class.
We obtain maximum likelihood estimates of all parameters and introduce a special numerical trick for calculating the maximum likelihood estimates of
the true number of consumers in each class (see Theorem \ref{Theorem}).

Details of our model are in Section \ref{sec:model}, with estimates
described in Section \ref{sec:estimation}.  In Section \ref{sec:apply}
we give details of implementation as well as extend our model and
estimation procedure to the case where there are replicate
observations from each transformer.  Section \ref{sec:data_analysis}
contains the analysis of the energy consumption data for the three
transformers. Section \ref{sec:simulation} contains the results of a
simulation study.

The estimation of typical load curves of
  electrical consumption using aggregated functional data was first
  done by  \cite{dias:garcia:martarelli:2009} and revisited in the
  Bayesian framework by \cite{dias:garcia:schmidt:2013}.   However, these authors assumed that consumers reported their true classes.

In summary, we develop and study statistical methodology for the analysis of curve data, where the distribution of each curve depends on a class membership variable. We do not observe data from individual curves, but only observe pointwise sums. Moreover, we do not know the exact counts for class membership, only approximate counts.  Our goal is to use the approximate counts and the aggregated curve data to
estimate the true number of individuals in each class and to estimate the typical curve of each class.   In addition, we estimate variance and covariance parameters.

  In the signal processing literature the problem of aggregated
  information is known as linear Blind Signal Separation (BSS). The
  simplest model assumes the existence of $C$ independent signals
  $W_1(\cdot), \ldots, W_C(\cdot)$ and the observation of at least $I$
  mixtures $Y_1(\cdot), \ldots, Y_I(\cdot)$, these mixtures being
  $Y_{i}(\cdot)=\sum_{c=1}^C a_{ic} W_c(\cdot)$, for $i=1,\ldots,I$ with
  unknown coefficients $a_{i1},\ldots,a_{iC}$. However, the BSS problem lies in the
  identification of the signals $W_1(\cdot), \ldots, W_C(\cdot)$ using only
  the observed data whereas in this paper we
  are interested in the mean and covariance functions of such
  processes. Usually the BSS methods are based on multivariate
  techniques (component analysis, orthogonalization, spatio-temporal
  decorrelation) which consider the time as a discrete set and do not
  take into account that the sources and the observations are
  continuous curves. For a review on the algorithms and the
  statistical principles of BSS see e.g. \cite{cardoso:1998},
  \cite{choi:cich:park:lee:2005}, and \cite{comon:jutten:2010}.
\\

\section{Notation and model}
\label{sec:model}

We index transformer by $i=1,\ldots, I$, class by $c=1,\ldots, C$ and  consumer served by transformer $i$ by $l=1,\ldots, N_i$.
We denote the true class of individual $l$ served by transformer $i$ as
$c_{li}$  and the reported class as $r_{li}$.
We let $W_{li}( t)$ equal the electricity consumption at time $t$ of individual $l$ served by transformer $i$.
 We model $W_{li}$ as a hidden, random process whose
distribution depends on  the value of $c_{li}$.
In transformer $i$, we do not observe the $W_{li}$'s but rather their  sum plus  measurement error,
at $n_i$ time points, $t_{i1} < t_{i2}< \cdots <t_{in_i}$.  For simplicity of notation and exposition, we only consider the case with $n_i \equiv n$ and
$ t_{ij} \equiv t_j$,  for all $i$ and for $j=1,\ldots,n$.
We do not observe $M_{ci}$, the true number of consumers in class $c$ served by transformer $i$.  Rather, we observe $R_{ci}$,
 the number
of reported consumers in class $c$ served by transformer $i$.
Throughout, we assume that random quantities are independent from transformer to transformer.

We are interested in estimating the true counts of consumers in each class, served by each transformer,
and the typical usage curve of a consumer of class $c$, which is simply
 the expected value of $W_{li}(t)$ when $c_{li}=c$.

%%%%%%%%%%%%%%%%%%%%%%%%%%%%%%%%%%%%%%%%%%%%%%%%%%%%%%%

\subsection{Model for $W_{li}$ and the observed aggregated electricity consumption}\label{sec:modelWli}

We suppose that,  for an individual of class $c$, the energy consumption is a stochastic process with distribution  depending on $c$ but
not on the transformer.   Specifically, we suppose that $W_{li}$, the energy consumption of consumer $l$ served by transformer $i$ of consumer type $c_{li}=c$,   is
given by
\[
W_{li}(t)  \bigg{|}_{c_{li}=c} =  ~~\alpha_c(t) + \alpha_{cli}^*(t)\]
where $\alpha_c$ is the  non-random typical usage curve (also called the typology) in class $c$
and  $\alpha_{cli}^*$ is the consumer-specific random perturbation. Using the notation that the $k$th component of a vector $\vbold$ is $\vbold[k]$,
we model $\alpha_c$ and $\alpha_{cli}^*$ with B-splines
basis functions $\phi_1,\ldots,\phi_K$ and $\psi_1,\ldots, \psi_{K^*}$.  Letting
$\bgamma(t)=(\gamma_1(t),\ldots,\gamma_K(t))'$ and $\bpsi(t)$ defined similarly,
\begin{equation}
\alpha_c(t) =   \sum_{k=1}^{K}  \bgamma^c[k] \phi_{k}(t)  \equiv \bphi(t)' \bgamma^c
\label{eq:alpha}
\end{equation}
and
\begin{equation}
\alpha_{cli}^*(t) = \sum_{k=1}^{K^*}  \bgamma^{cli}[k] \psi_k(t)  \equiv \bpsi(t)'\bgamma^{cli}.
\label{eq:alphastar}
\end{equation}

We suppose that
the vector $\bgamma^c$ is an unknown parameter vector and the vectors $\bgamma^{cli}$ are random effects,
normally distributed, independent
with mean 0 and covariance matrix $\Sigma_c$.
Note that this implies that, given $c_{li}=c$,  $W_{li}$ is a Gaussian process with  mean $\alpha_c$ and covariance function
\[
\sigma_c(s,t)
\equiv
\cov(W_{li}(s), W_{li}(t))
= \cov(\alpha_{cli}^*(s), \alpha_{cli}^*(t)) =  \bpsi(s)'  ~\Sigma_c ~\bpsi(t).
\]
The process $\alpha_{cli}^*$ allows us
to account for within consumer correlation over time.  We assume that the $\alpha_{cli}^*$'s
are independent processes.

Notice that the vectors of basis function evaluations $\bphi(t)$ and $\bpsi(t)$ in (\ref{eq:alpha}) and (\ref{eq:alphastar}), respectively, do not depend on the consumer type $c$ -- that is, we use the same size of basis ($K$ and $K^*$) and the same knot locations for all types of
consumers. One could consider a different number of basis functions and  different knot locations for each consumer class. For instance, one could place more knots around 6-9pm for residential consumers as they show a spike in electricity consumption during this time of the day. We also use
the same basis functions for each transformer,  that is $\bphi(t)$ and $\bpsi(t)$ in (\ref{eq:alpha}) and (\ref{eq:alphastar}), respectively, do not depend on $i$.

It is  important to note that, in our model,  $\alpha_c$ cannot depend on the transformer $i$, that is, that the $\bgamma^c$'s in
(\ref{eq:alpha}) can not depend on $i$.   If  the $\bgamma^c$'s were to depend on $i$, we would be
unable to estimate them from our data:  our data consist of  just  one total energy curve per transformer, a curve that is a weighted sum of
the unknown class-level total energy  curves.   By using the same $\bgamma^c$'s for all transformers, we are able to pool information across all of the transformers' total energy curves in order to estimate the $\bgamma^c$'s.

Choosing the number of basis functions $K$, which is equivalent to
choosing the number of knots, has been a subject of great research interest. Several authors suggested algorithms in
order to provide a good choice of the dimension of the approximant
space as a function of the sample size; see, for example,
\cite{gu:1993b}, \cite{anto:1994}, \cite{bodi:vill:lars:wahl:2000},
\cite{kohn:marr:yau:2000} and \cite{devo:petro:teml:2003}. However,
all of these procedures, including adaptive ones such as
\cite{koop:ston:1991}, \cite{luo:wahb:1997} and \cite{dias:1998a}, deal with
a non-random choice of $K$.

For consumers served by transformer $i$, we observe the total energy
use plus error.  We assume that the magnitude of the variability of
the error does not depend on the transformer.  That is, we observe the
data vector $\vecY_i \equiv (Y_i(t_{i1}),\ldots,Y_i(t_{in}))'$ where,
for $\epsilon_i(\cdot)$ and $W_{li}(\cdot)$ independent, with
$\epsilon_i$ Gaussian white noise with var($\epsilon_i(t))= \sigma^2$,
 \begin{eqnarray}
 Y_i(t) &=& \sum_{l=1}^{N_i}  \sum_{c=1}^C W_{li}(t)  ~\I\{ c_{li}=c\}
 ~ + ~ \epsilon_i(t)
 \nonumber\\
 &=&
  \sum_{c=1}^C  M_{ci} ~\alpha_c(t) ~+~   \sum_{l=1}^{N_i}  \sum_{c=1}^C \alpha^*_{cli}(t)\I\{ c_{li}=c\}
 ~ + ~ \epsilon_i(t) .
 \label{eq:Yit}
  \end{eqnarray}
  We easily see that
 E$(Y_i(t)) =  \sum_{c=1}^C  M_{ci} ~\alpha_c(t) $
and, because of the independence of the $\alpha_{cli}^*$'s,
\[
\cov(Y_i(s),Y_i(t)) =
\sum_{l=1}^{N_i}  \sum_{c=1}^C  \I \{ c_{li}=c\} ~ \cov( \alpha_{cli}^*(s), \alpha_{cli}^*(t))
~+~ \cov(\epsilon_i(s),\epsilon_i(t)).
\]
 Defining the $n$ by $K$ matrices $\Phi$ and $\Psi$ as
 \[
\Phi[j,k] = \phi_k(t_{ij})  {~~\rm{ and ~~}}  \Psi[j,k]  = \psi_k(t_{ij})
\]
yields
\begin{equation}
\label{eq:Yi.normal}
\vecY_i \sim {\rm{N}}\left( \Phi \sum_{c=1}^C M_{ci}  \bgamma^c,
\Psi \sum_{c=1}^C M_{ci} \Sigma_c \Psi ' + \sigma^2 \I\right).
\end{equation}

From (\ref{eq:Yi.normal}), we see that, without further information about the $M_{ci}$'s, the distribution of $\vecY_i$ is not identifiable - there are an infinite number of distinct $M_{ci}$'s and $\gamma^c$'s yielding the same distribution. However, when we observe the reported counts, the joint distribution of $\mathbf{Y}_i$ and $R_{1i},\dots,R_{Ci}$ is identifiable, provided we know the rates of misreporting.

Equation (\ref{eq:Yit}) considers the case where there is only one $W_{li}$
for each consumer and can be used to model the situation where
$W_{li}$ is the power usage on a specified date or the average of usages on a sequence of dates. We can easily extend (\ref{eq:Yit}) to model the situation where there are $D_{i}$ replicates of the $W_{li}$'s for consumer $l$ served by
transformer $i$. We use this extension in the simulation study
and in the data analysis to model a consumer's energy consumption on a
sequence of five days, considering these five days as five replicates.
The calculation and maximization of the likelihood for the replicate
case is a straightforward modification of the iterative procedure described in
Section \ref{sec:estimation}.

%%%%%%%%%%%%%%%%%%%%%%%%%%%%%%%%%%%%%%%%%%%%%%%%%%%%%
\subsection{Model for the reported counts of consumer classes}
\label{sec:model.Ri}

Recall that $r_{li}$ is the class reported by consumer $l$ served by transformer $i$,  that $c_{li}$ is the consumer's true class,  that
$M_{ci}$ is the true number of consumers of class $c$ served by transformer $i$ and that  $R_{ci}$ is the corresponding reported number.  We suppose that all
consumers report some class, that is, that $N_i$, the total number of consumers served by transformer $i$, is equal to
$ \sum_c R_{ci} $, which is also equal to $ \sum_c M_{ci}$.  Recall that, in our model, the $M_{ci}$'s are fixed parameters and the $R_{ci}$'s are random variables.  We require a model for consumer reporting, that is, for $R_{1i},\ldots,R_{Ci}$ for true counts $M_{1i},\ldots, M_{Ci}$.

To model consumer reporting, we suppose that there is a known {\em{fraud matrix}} $\Fraud$, not depending on the transformer, with
\[
\Fraud(c,r) = P\{ r_{li}=r |  c_{li}=c \}, ~ r, c=1,\ldots,C,
\]
the probability that a consumer of class $c$ reports as being of class $r$.
We assume that consumers report independently of other consumers.

Table \ref{eq:Xmat}, a  $C$ by $C$ table of counts,
 is useful in understanding the distribution of the reported counts.  For convenience, we drop the transformer subscript $i$.
Let $x_{cj}$ be the number of consumers who are of class $c$ but report they are of class $j$.
The reported number of consumers of class $j$  is the sum of counts in column $j$:
$R_{j} = \sum_c  x_{cj}$.
The true number of consumers of class $c$  is the sum of counts in row $c$: $M_{c} = \sum_j  x_{cj}.$

For $c=1,\ldots,C$, let
$\vecX_{c} \equiv (x_{c1},\ldots, x_{cC})'$, the vector of reported counts from consumers of class $c$,  found
in row $c$ of the count table with row total equal to $M_c$.
 Then $\vecX_{c}$ is multinomial$(M_{c}, \Fraud(c,1),\ldots, \Fraud(c,C))$.  By independence of consumer reporting,
 $\vecX_{1},\ldots,\vecX_{C}$
are  independent.  Thus, we have  defined the joint distribution  of $R_1=\sum_c  x_{c1},\ldots, R_C
=\sum_c  x_{cC}$ when the true class counts are $M_1,\ldots, M_C$.

%%%%%%%%%%%%%%%%%%%%%%%%%%%%%%%%%%%%%%%%%%%%%%%%%%%%%%%

\section{Maximum Likelihood Estimation}
\label{sec:estimation}

We estimate the parameters by maximizing the log likelihood.  Let the parameter vectors describing the $W_{li}$ processes be those defining the class means,
the $\alpha_c$'s, that is, let the set of parameter vectors describing the $W_{li}$ processes be
$\calG=  (\bgamma^1,\ldots,\bgamma^C)'$.  Let the set of parameters  defining the class variance structure be $\calS = \{ \Sigma_1,\ldots,\Sigma_C\}$.   Recall that $\sigma^2$ is the
measurement error variance.
Let $\vecM_i = (M_{1i},\ldots, M_{Ci})'$ be  the vector of true class counts in transformer $i$
and $\vecM$  the collection of true counts, $\vecM_1,\ldots,\vecM_I$.  Recall that the elements of $\vecM$ are unknown parameters.

The data are $\vecY_i$ as in (\ref{eq:Yi.normal}) and   $\vecR_i = (R_{1i},\ldots, R_{Ci})'$,  the reported counts, for transformers $i=1,\ldots, I$.

By the   independence of the transformers, the log likelihood is
\[{\cal{L}}(\calG,\calS, \vecsigma^2,\vecM) = \sum_{i=1}^I  {\cal{L}}_i(\calG,\calS, \sigma^2,
 \vecM_i)
\]
where
 \begin{eqnarray}
 \label{eq:el.i}
{\cal{L}}_i(\calG,\calS, \sigma^2,
 \vecM_i)& \equiv&
  {\cal{L}}_i( \calG, \calS, \sigma^2, \vecM_i \big{|} \vecY_i, \vecR_i ) \nonumber
 \\ &=&
 \log ~\big[ ~f_{Y_i}(\vecY_i \big{|} \calG, \calS, \sigma^2, \vecM_i)  \times P\{\vecR_i |\vecM_i\}~\big]
\end{eqnarray}
where $ P\{\vecR_i \big{|} \vecM_i\}$ is the probability mass function of $\vecR_i$ calculated assuming that the true consumer category counts for transformer $i$ are the components of  the  vector  $\vecM_i$.
The expression for $\log[ f_{Y_i}(\vecY_i  \big{|} \calG,\calS, \sigma^2, \vecM_i) ]$ follows directly from
 our model (\ref{eq:Yi.normal}).
The other part  of ${\cal{L}}_i$, $ P\{\vecR_i \big{|} \vecM_i\}$, is discussed in Sections  \ref{sec:R.given.M.in.L}
and \ref{sec:R.given.M}.

%%%%%%%%%%%%%%%%%%%%%%%%%%%%%%%%%%%%%%%%%%%%%%%%%%%%%
\subsection{Maximization procedure}
\label{sec:maximization.procedure}

We carry out the maximization of ${\cal{L}}$ iteratively and step-wise, where, at the $s$th iteration, we update the parameter estimates
$\calG^{(s)}$,  $\calS^{(s)}$, $\vecsigma^{2(s)}$ and $\vecM^{(s)}$    to
   $\calG^{(s+1)}$,  $\calS^{(s+1)}$,    $\vecsigma^{2(s+1)}$ and $\vecM^{(s+1)}$ so that
${\cal{L}}(\calG^{(s+1)},  \calS^{(s+1)}, \vecsigma^{2(s+1)},
\vecM^{(s+1)}) $
$\ge
{\cal{L}}(\calG^{(s)},  \calS^{(s)}, \vecsigma^{2(s)},$ $
\vecM^{(s)}) $.
We initialize the procedure by taking  $\vecM_i^{(0)} \equiv \vecR_i$.  We then carry out the following
two steps until convergence.  Details of each step follow in Sections \ref{sec:update}, \ref{sec:R.given.M.in.L} and \ref{sec:apply}.

\begin{enumerate}
\item Given $\vecM^{(s)}$, we let $\calG^{(s+1)}$, $\calS^{(s+1)}$ and $\vecsigma^{2(s+1)}$
maximize the log likelihood, or at least not decrease the log likelihood.
 \item  Given $\calG^{(s+1)}$, $\calS^{(s+1)}$  and $\vecsigma^{2(s+1)}$,  we
 let $\vecM^{(s+1)}$
maximize the log likelihood, or at least not decrease the log likelihood.
\end{enumerate}

\vskip 20pt

 We have no theory to prove that the maximum likelihood is unique or that our iterative procedure converges.   However, in all of our analyses - of simulated data and of the transformer data - our procedure always converged.  To study the possible problem of multi-modality of the likelihood function, for a few data sets we used several different sets of starting values for the parameter estimates.  In all cases, the algorithm converged to the same final parameter estimates.

 \vskip 10pt

 %%%%%%%%%%%%%%%%%%%%%%%%%%%%%%%%%%%%%%%%%%%%%%%%%

 \subsection{Step 1: updating  $\calG$, $\calS$ and $\vecsigma^2$
 }
\label{sec:update}

Using $\vecY_i$'s normal distribution in
(\ref{eq:Yi.normal}), we see that we must minimize
\[
l_1 (\calG,\calS,\vecsigma^2)
\equiv \sum_{i=1}^I  \log \Big{|}  \Psi \sum_{c=1}^C M_{ci} \Sigma_c \Psi' + \sigma^2\I \Big{|}
\]
\[
+
\sum_{i=1}^I
 \left(\vecY_i  -  \sum_{c=1}^C M_{ci} \Phi \bgamma^c \right)'
  \left( \Psi \sum_{c=1}^C M_{ci} \Sigma_c \Psi' + \sigma^2 \I   \right)^{-1}
    \left(\vecY_i -  \sum_{c=1}^C  M_{ci} \Phi \bgamma^c\ \right)
    \]
    as a function of  $\calG $,  $\calS $  and
    $\vecsigma^2 $, keeping the $M_{ci}$'s fixed.
    We carry this out iteratively, in three steps:

\begin{enumerate}
\item
\label{Step1}
Given  ${\calS}^{(s)}$  and  $\vecsigma^{2(s)}$, let $\calG^{(s+1)}$ minimize $l_1$.
This step poses no problem and can be done explicitly yielding
\begin{eqnarray}
\calG^{(s+1)} &=& \left(\displaystyle\sum_{i=1}^{I} \left[M_{1i}^{(s)}\Phi \cdots M_{Ci}^{(s)}\Phi \right]' \Lambda_i^{-1} \left[ M_{1i}^{(s)}\Phi \cdots M_{Ci}^{(s)}\Phi \right] \right)^{-1} \nonumber \\
& & \quad \times
\left( \displaystyle \sum_{i=1}^{I} \left[M_{1i}^{(s)}\Phi \cdots M_{Ci}^{(s)}\Phi \right]' \Lambda_{i}^{-1} \mathbf{Y}_i \right), \label{eq:1a}
\end{eqnarray}
where $\Lambda_i = \left(\Psi\sum_{c=1}^{C}M_{ci}^{(s)}\Sigma_c^{(s)} \Psi' + \sigma^{2(s)} \mathbf{I}\right)$.

\item
\label{Step2}
Given $\calG^{(s+1)}$ and ${\calS}^{(s)}$, find
  $\vecsigma^{2(s+1)}$ that minimizes $l_1$. This would generally
  require a numerical minimization.
\item
\label{Step3}
Given $\calG^{(s+1)}$ and $\vecsigma^{2(s+1)}$, let
  $\calS^{(s+1)}$ minimize $l_1$.  This is the most challenging step
  and will typically require simplifying assumptions of the form of
  the $\Sigma_c$'s, as in Section \ref{sec:apply}.  In addition, this
  would generally require a numerical minimization.
\end{enumerate}

%%%%%%%%%%%%%%%%%%%%%%%%%%%%%%%%%%%%%%%%%%%%%%%%%

\subsection{Step \ref{Step2}:  updating $\vecM_1,\ldots,\vecM_I$}
\label{sec:R.given.M.in.L}

  Note that, to maximize the log likelihood with respect to $\vecM_1,\ldots, \vecM_I$, we can carry out $I$ separate maximizations,
  one for each transformer.  That is, for each fixed $i=1,\ldots,I$, we seek $M_{ci}$, $c=1,\ldots,C$ to
 maximize  ${\cal{L}}_i(\calG,\calS, \sigma^2,
 \vecM_i) $ in (\ref{eq:el.i}), treating $\calG$, $\calS$, and $ \sigma^2$ as fixed.

This step brings challenges.  The function  $P\{ \vecR_i \big{|} \vecM_i\}$ does not have a closed form, nor does its derivative.
Furthermore, Newton-Raphson type methods of maximization are inappropriate since each  $M_{ci}$ is an integer with plausible values usually within a small range of the reported
count $R_{ci}$.  Therefore the possible values of $M_{ci}$ must be treated as integer.

We can approximate the function  $P\{ \vecR_i \big{|} \vecM_i\}$ via simulation.
The most natural simulation is a ``brute force" one:  for each fixed $\vecM_i$ we would generate a large number of $\vecR_i$'s according to the model described in
Section \ref{sec:model.Ri} and calculate the proportion of times the generated $\vecR_i$ is equal to the observed $\vecR_i$.   For instance, consider the case that transformer $i$ serves 50 consumers, of two classes, with the reported number of residential consumers  $R_{1i}=$40 and the reported number of commercial consumers $R_{2i}=$10.   We would want to calculate
$P\{R_{1i}=40, R_{2i} =10 | M_{1i}=m_1, M_{2i}=50-m_1\}$  for all values of $m_1$, or at least all plausible values of $m_1$.   Thus, we would want to carry out many simulations -- 51 if we wanted to consider all possible values of $m_1$.  We would have to do this for each transformer.    Clearly, a short cut would be
desirable.

 Fortunately, we have determined a less computer intensive simulation method to approximate
$P\{ \vecR_i \big{|}  \vecM_i\}$,
justified by Theorem \ref{Theorem}  in the next section.  We show that we can calculate
$ P\{ \vecR_i \big{|} \vecM_i\}$ in ${\cal{L}}_i$ via a function $H_i$.  We can   easily and quickly approximate and  table  $H_i(m)$ for
all values of $m$ with only one simulation per transformer.
  Thus, to maximize  ${\cal{L}}_i$, by Theorem \ref{Theorem}, it
 suffices to find $\vecM_i$ to maximize
\[
{\cal{L}}_i^*(\vecM_i | \vecY_i, \vecR_i) \equiv {\cal{L}}_i^*(\vecM_i ) \equiv
 -  \frac{1}{2}\log\Big{|} \Psi \sum_{c=1}^C M_{ci} \Sigma_c \Psi' + \sigma^2\I \Big{|}
\]
\[
 -\frac{1}{2}
 \left(\vecY_i  -  \sum_{c=1}^C M_{ci} \Phi \bgamma^c \right)'
  \left( \Psi \sum_{c=1}^C M_{ci} \Sigma_c \Psi' + \sigma^2 \I   \right)^{-1}
    \left(\vecY_i -  \sum_{c=1}^C M_{ci} \Phi \bgamma^c\ \right)
    \]
    \[
    +
  \sum_{c=1}^C \log  M_{ci}!   + \log H_i(\vecM_i).
\]

In our data analyses and simulation studies, we easily calculated
${\cal{L}}_i^*(\vecM_i) $ for all possible values of $\vecM_i$ and chose the value of $\vecM_i$ that maximized ${\cal{L}}_i^*$.   If this is not practical,
then one could calculate ${\cal{L}}_i^*$ for a small range of values of $\vecM_i$.

%%%%%%%%%%%%%%%%%%%%%%%%%%%%%%%%%%%%%%%%%%%%%%%%%%%%%

\subsection{Alternate form for P$(\vecR_i | \vecM_i)$}
\label{sec:R.given.M}

For convenience, we will once again drop $i$, the subscript indicating the transformer.
  Recall the notation and the definitions in Section \ref{sec:model.Ri}, where
we defined the joint distribution of $R_1,\ldots,R_C$ by defining the distribution  of the  rows of Table
\ref{eq:Xmat}.   Here,
we derive an expression for the probability that  $R_1=\rval_1,\ldots, R_C=\rval_C$ in terms of  random vectors,
$\tilde{\vecX}_1,\ldots,\tilde{\vecX}_C$,  that resemble
  the columns of Table \ref{eq:Xmat}.   Specifically, for $j=1,\ldots,C$, let  $\tilde{\vecX}_j$ be multinomial with
parameters $\rval_j,
p_{1j},p_{2j},  \ldots ,$ and $p_{Cj}$ with
 \begin{equation}
p_{cj} = \frac{ \Fraud(c,j)}{ \sum_{l=1}^C \Fraud(l,j)}.
\label{eq:pcj}
\end{equation}
  Thus, the  entries of each $\tilde{\vecX}_j$ sum to $\rval_j$ and the multinomial probability associated with the $c$th component of $\tilde{\vecX}_j$ is proportional to $\Fraud(c,j)$,
the probability that a consumer of class $c$ reports being in class $j$.  That is, $\tilde{\vecX}_j$ is a vector of counts, dividing up the $\rval_j$ consumers
 who have reported
being class $j$ into their true classes. While the distribution of  $\tilde{\vecX}_j$ is not equal
to the conditional distribution of column $j$ of Table \ref{eq:Xmat} given  $R_j=\rval_j$,  the distribution of   $\tilde{\vecX}_j$ does relate to
the distribution of $R_1,\ldots, R_C$, as given in the following theorem.

\begin{theorem}
\label{Theorem}
For the random variables $R_1,\ldots,R_C$ as defined in Section \ref{sec:model.Ri} and for $\tilde{\vecX}_1, \ldots, \tilde{\vecX}_C$ independent with
distributions as defined above,
\begin{eqnarray}
P\{ \vecR = \vecr | \vecM = \vecm\} &=& P \{R_1 = \rval_1, \ldots, R_C=\rval_C| M_1=m_1 \ldots, M_C= m_C\}
\nonumber \\
&=&
  \frac{\prod_{j=1}^C \left[ \sum_{c=1}^C \Fraud(c,j) \right]^{r_j}}{\prod_{j=1}^C \rval_j!} \times  \prod_{c=1}^C m_c!
 \times  H(m_1,\ldots,m_C)
\nonumber \end{eqnarray}
where
\begin{equation}
H(m_1,\ldots,m_C) =
{\rm{E}} \left( {\rm{I}} \bigg{\{}  \sum_{j=1}^C \tilde{\vecX}_j[c] = m_c, c=1,\ldots, C\bigg{\}} \right).
\label{eq:Hfunction}
\end{equation}
\end{theorem}

\vskip 10pt

\noindent{\em{Comment}}. Theorem \ref{Theorem} allows us to approximate  $P \{\vecR=\vecr | \vecM=\vecm \}$
for fixed $\vecr$
via one simulation study.
To understand this, consider the following simple example.  Suppose there are
two classes and we have data from a transformer that serves  75 clients,
with reported counts $\rval_1=32$ and $\rval_2=43$.
Suppose that the fraud matrix is
\begin{equation}
\Fraud =
\left[
\begin{array}{ccc}
0.98  &  0.02
\\
0.05  &  0.95
\end{array}
\right].
\label{eq:Fraud}
\end{equation}
The ``brute force" way  to approximate $P\{ R_1 = 32, R_2=43 | M_1 = m_1, M_2 = m_2\}$ (the way we avoid) is to carry out one simulation study  for each vector $(m_1, m_2)$, generating vectors
$\vecX_1,\vecX_2$ and counting the proportion of times that
$\vecX_1 + \vecX_2 $ equals (32,43)$'$.   Fortunately,
the Theorem gives
a form for $P\{ R_1 = 32, R_2=43 | M_1 = m_1, M_2 = m_2\}$  that allows us to carry out just one simulation study that can then be used for all vectors $(m_1, m_2)$, as follows.    We simulate $B$ data sets, with the $b$th simulated data set
consisting of two independent multinomial vectors:
\[
\tilde{\vecX}_1^b  ~\sim ~{\rm{~ multinomial}}\left(32,\frac{0.98}{1.03}, \frac{0.05}{1.03},\right),
\]
and
\[
\tilde{\vecX}_2^b ~\sim ~{\rm{~ multinomial}}\left(43,\frac{0.02}{0.97}, \frac{0.95}{0.97}\right).
\]
Let $M_1^b$ and $M_2^b$  correspond to the row totals:
$M_c^b =  \tilde{\vecX}_1^b[c] +  \tilde{\vecX}_2^b[c] $.
We then approximate ${H}$  via $\hat{H}$,
using the $M_1^b$'s and $M_2^b$'s  that result from the simulation.
For instance, $\hat{H}(34,42)$ is the proportion of the $B$ simulated data sets that had $M_1^b=34$ and $M_2^b=42$.

We carry out this procedure with $B=100,000$ runs.  Table \ref{table:m1.example} shows the resulting approximations of $H(m_1, 75-m_1)$ when
 $R_1=32$ and $R_2=43$.

In general, to estimate $H$, we simulate $B$ independent sets distributed as $\{ \tilde{\vecX}_1, \ldots,  \tilde{\vecX}_C\}$, with the $b$th simulated
data set denoted
$\{ \tilde{\vecX}_{1b}, \ldots,  \tilde{\vecX}_{Cb}\}$, $b=1,\ldots,B$.  We let
\[
\hat{H}(m_1,\ldots, m_C)  = \frac{1}{B} \sum_{b=1}^B
{\rm{I}} \bigg{\{}  \sum_{j=1}^C \tilde{\vecX}_{jb}[c] = m_c, c=1,\ldots, C\bigg{\}}.
\]

\vskip 15pt
\noindent
{\em{Proof of Theorem 1}}.
Write
$P \{\vecR=\vecr|  \vecM = \vecm \}$ as
\[ \sum_{x_{cj}: \sum_c x_{cj} = \rval_j}
P\{ \vecX_1=(x_{11}, \ldots, x_{1C})\}~\times \]
\[\times~
P\{ \vecX_2=(x_{21}, \ldots, x_{2C} )\}   \times \cdots \times
P\{ \vecX_C=(x_{C1}, \ldots, x_{CC})\}
\]
\[ = \sum_{ \substack{ x_{cj}: \sum_c x_{cj} = \rval_j \\ {\rm{and}}~\sum_j x_{cj} = m_c}}
\Bigg{\{}  \left[ \frac{ m_1!}{ x_{11}! \cdots x_{1C}!}  \Fraud(1,1)^{x_{11}}  \cdots \Fraud(1,C)^{x_{1C}}  \right] \times \cdots
\]
\[
\times
\left[ \frac{ m_C!}{ x_{C1}! \cdots x_{CC}!}  \Fraud(C,1)^{x_{C1}}  \cdots \Fraud(C,C)^{x_{CC}} \right]
\Bigg{\}}.
\]
Rearranging terms yields that the probability is equal to
\[
\frac{ \prod_{c=1}^Cm_c!} { \prod_{c=1}^C\rval_c!} ~
 \sum_{ \substack{ x_{cj}: \sum_c x_{cj} = \rval_j \\ {\rm{and}}~\sum_j x_{cj} = m_c}}
 \Bigg{\{}
\left[ \frac{ \rval_1!}{ x_{11}! x_{21}!   \cdots x_{C1}!}
\Fraud(1,1)^{x_{11}}  \cdots \Fraud(C,1)^{x_{C1} }\right]
\times \cdots
\]
\[\times
\left[ \frac{ \rval_C!}{ x_{1C}! x_{2C}!   \cdots x_{CC}!}
\Fraud(1,C)^{x_{1C}}  \cdots \Fraud(C,C)^{x_{CC} }\right]
\Bigg{\}}
\]
\begin{eqnarray}
&=&
\frac{ \prod_{c=1}^Cm_c!} { \prod_{c=1}^C\rval_c!} ~
\sum_{ \substack{ x_{cj}: \sum_c x_{cj} = \rval_j \\ {\rm{and}}~\sum_j x_{cj} = m_c}}
\left[ \frac{ \rval_1!}{ x_{11}! x_{21}!   \cdots x_{C1}!}
p_{11}^{x_{11}}   p_{21}^{x_{21}}
\cdots p_{C1}^{x_{C1}}  \right]
\times  \cdots
\nonumber \\
&&
\times \left[\frac{ \rval_C!}{ x_{1C}! x_{2C}!   \cdots x_{CC}!}
p_{1C}^{x_{1C}}   p_{2C}^{x_{2C}}     \cdots p_{CC}^{x_{CC}}   \right]
~~\times \prod_{j=1}^C \left[ \sum_{c=1}^C \Fraud(c,j) \right]^{\rval_j}
\nonumber \\
&=&
\frac{\prod_{j=1}^C \left[ \sum_{c=1}^C \Fraud(c,j) \right]^{\rval_j}}
{\prod_{c=1}^C\rval_c!  }
 \times  \prod_{c=1}^C m_c! \times \nonumber  \\
&& ~~ \times
\sum_{  x_{cj}: \sum_j x_{cj} = m_c}
{\prod_{j=1}^C   }
P\{ \tilde{\vecX}_j = (x_{1j},\ldots,x_{Cj})\}
\nonumber \\
&=&
\frac{\prod_{j=1}^C \left[ \sum_{c=1}^C \Fraud(c,j) \right]^{\rval_j}}
{\prod_{c=1}^C\rval_c!  }
 \times  \prod_{c=1}^C m_c!
\times
H(m_1,\ldots,m_C).
\nonumber
\end{eqnarray}

%%%%%%%%%%%%%%%%%%%%%%%%%%%%%%%%%%%%%%%%%%%%%%%%%%%%%
\section{Details of Implementation}
\label{sec:apply}

We now give  details  of implementation concerning  choice of starting values of the parameter estimates,  maximization of the likelihood under the
restriction that  $\Sigma_c = \sigma^{2}_{\gamma,c} \I$ and the choice of software.

As previously mentioned, we recommend using  the reported counts as the initial estimates of the true counts.

Recall from Section \ref{sec:modelWli} that we assume the magnitude of the variability of the error
does not depend on the transformer, that is, $\sigma^2_i=\sigma^2$ for all $i=1,\ldots,I$. To obtain the initial estimate of $\sigma^2$ in the replicate case, we first fit a smoothing spline curve to each replicate in transformer $i$ and calculate the sample variance of the residuals of that fit adjusting for the correct degrees of freedom, which are based on the trace of the smoothing hat matrix. We then pool those variances across replicates and transformers to obtain our initial estimate of $\sigma^2$. For the non-replicate case the procedure is the same, but we only need to pool across transformers. More details on smoothing based estimation of variances and calculation of appropriate degrees of freedom can be found in  \citet{wahba1983bayesian}.

Since we assume that the $\bgamma^{cli}$'s  have covariance
$\Sigma_c= \sigma_{\gamma,c}^2 \I$,  the set of covariance parameters
$\calS$ is equal to $  \{ \sigma_{\gamma,c}^2, c=1,\ldots,C\}$.
We use method of moments for our initial estimates of the
  $\sigma_{\gamma,c}^2$'s, using the fact that $\vecY_i $ has a multivariate distribution with   covariance
  matrix
  $ [\sum_{c=1}^C  M_{ci} \sigma_{\gamma,c}^2 ]   \Psi \Psi'  + \sigma^2\I] $.   For the purpose of calculating
  the  initial estimates,
  we suppose that prior knowledge tells us that
  $\sigma_{\gamma,c}^2 = s_c  \sigma_{\gamma,C}^2$ for some known $s_c$'s, $c=1,\ldots, C-1$.  To extend this notation, we set $s_C =1$.

 In the non-replicate case, we write
  \begin{equation}
  \sum_{i , j}  \var( \vecY_i[j])   =
\sigma_{\gamma,C}^2  \sum_i  \left[\sum_{c=1}^C  M_{ci} s_c  \right]  {\rm{trace}}( \Psi \Psi')  +  I n \sigma^2.
   \label{eq:sigma_MOM}
   \end{equation}
  We use our initial estimates of $M_{ci}$ and
    $\bgamma^c$  to form the estimate of the left side of (\ref{eq:sigma_MOM}):
    \[
    \sum_j \widehat{ \var}( \vecY_i[j])  = || \vecY_i - \widehat{M}_{ci}^{(0)}  \Phi   \hat{ \bgamma}^{c(0)} ||^2.
    \]
  We
  substitute our
     estimates of the $M_{ci}$'s and  the $\sigma^2$'s into the right side of (\ref{eq:sigma_MOM}) and then solve for our estimate of
   $\sigma_{\gamma,C}^2$.
     This  yields our initial estimate,
     $\hat{\sigma}_{\gamma,C}^{2(0)}$, and our other initial estimates, $\hat{\sigma}_{\gamma,c}^{2(0)}  \equiv s_c \hat{\sigma}_{\gamma,C}^{2(0)}$.

   In the case where we observe $D$ replicates from transformer $i$, we modify the calculations, summing both sides of
   (\ref{eq:sigma_MOM}):
   \[
   \sum_{d,i,j}  \var( \vecY_{i,d} [j])   =
D  \Bigg{\{}  \sigma_{\gamma,C}^2  \sum_i  \left[\sum_{c=1}^C  M_{ci} s_c  \right]  {\rm{trace}}( \Psi \Psi')  +  I n \sigma^2 \Bigg{\}}
% \label{eq:sigma_MOM_rep}
   \]
and estimating the variance of $\vecY_{i,d}[j]$ by the sample variance
   of $\vecY_{i,1}[j]$, $\ldots, \vecY_{i,D}[j]$.

To maximize the likelihood,  we carry out Steps \ref{Step1}, \ref{Step2} and \ref{Step3} of the updating algorithm of
Section \ref{sec:update}.  Step \ref{Step1} is straightforward.  For the one-dimensional maximization of Step \ref{Step2}, that is, for
updating estimates of $\sigma^2$,
we use the $R$ function {\it optimize}.

For
Step \ref{Step3} of  the updating algorithm, we must minimize $l_1$ with respect to the
$\sigma_{\gamma,c}^2$'s.
First, write
\[
   \Psi   \sum_{c=1}^C M_{ci} \Sigma_c   \Psi'  + \sigma^2 {\rm{I}}
   =
 \left(\sum_{c=1}^C  M_{ci} \sigma^{2}_{\gamma,c}\right)  \Psi     \Psi'  + \sigma^2 {\rm{I}} .
\]
Writing  the eigenvalue-eigenvector decomposition $\Psi \Psi'  $ as $Q' \Gamma Q $ with $\Gamma$ diagonal and $Q$ orthonormal yields
\[
\left[
 \left(\sum_{c=1}^C M_{ci} \sigma^{2}_{\gamma,c}\right)  \Psi     \Psi'  + \sigma^2 {\rm{I}} \right]^{-1}
 =
Q'  \left[
 \left(\sum_{c=1}^C M_{ci} \sigma^{2}_{\gamma,c}\right) \Gamma  + \sigma^2 {\rm{I}} \right]^{-1}
Q.
\]
Let  $\Delta_i = \Delta_i(\sigma^2_{\gamma,1}, \ldots, \sigma^2_{\gamma,C})$ be the diagonal matrix
\[
\Delta_i = \left(\sum_{c=1}^C M_{ci} \sigma^{2}_{\gamma,c}\right) \Gamma + \sigma^2 {\rm{I}}.
\]
Thus,  letting
$\vecY_i^* = Q ( \vecY_i - \sum_c M_{ci} \Phi \bgamma^c)$,
we must find $\sigma^2_{\gamma,1}, \ldots, \sigma^2_{\gamma,C}$ to
 minimize
\[l_1(\calG,\{\sigma^2_{\gamma,1}, \ldots, \sigma^2_{\gamma,C}\} ,\vecsigma ^2) =
 \sum_i \log \left| \Delta_i
\right|
 +
\sum_i  {\vecY_i^{*}}'\Delta_i^{-1} \vecY_i^*.
\]

 Since $\Delta_i$ is diagonal,   the minimization can be easily carried out numerically via a
  $C$-dimensional optimization.  Here, we have used the $R$ function {\it optim}  with method L-BFGS-B, which is a modification of the quasi-Newton method.  This function allows specification of a lower and upper bound for each variable, which we need to force variance parameter estimates to be non-negative.
Note that we can
    calculate $Q$ and $\Gamma$ at the beginning of our iterations, as
    they are determined by the choice of basis.
   The modification
    of Step \ref{Step3} for the replicate case is similar.

%%%%%%%%%%%%%%%%%%%%%%%%%%%%%%%%%%%%%%%%%%%%%%%%%%%%%
\section{Data analysis}
\label{sec:data_analysis}
Recall that our data set consists of energy consumption data  from three  transformers, recorded every 15 minutes
during five
days of a particular week. For each transformer, we have  $C=3$ consumer types, residential monophasic $(c=1)$, residential biphasic $(c=2)$ and commercial $(c=3)$.
In our analysis, we use the methods of Section {\ref{sec:apply} with $D=5$  replicates, one for each weekday.

In the analysis, we consider the fraud matrix given by
\begin{equation}
\Fraud =
\left[
\begin{array}{ccc}
0.96  &  0.02 & 0.02
\\
0  &  0.98 & 0.02 \\
0.05 & 0.05 & 0.9
\end{array}
\right],
\label{eq:Fraud_3classes}
\end{equation}
where a commercial consumer self-reports as either a monophasic or biphasic residential consumer, each
 with probability 0.05.  A residential
biphasic consumer never self-reports as monophasic and self-reports
as a commercial consumer with probability 0.02.  A monophasic consumer
self-reports as biphasic with probability 0.02 and as commercial with
probability 0.02. The misreporting occurs because commercial consumers
pay a higher rate for their energy than residential consumers and,
therefore, small business owners may report themselves as either
residential monophasic or biphasic consumers.  The other type of
misreporting is rare, but possible.  For example, a consumer with a
business and residence at the same location may close the business but
fail to notify the energy company. Another example would be a
residential consumer that self-reports biphasic but actually behaves
as a monophasic consumer, using only 127 volts. On the other hand, a
biphasic consumer cannot self-report as a monophasic consumer because
the energy company knows the voltage power of the residences.  All
this information can be obtained from the experts in the field.

We model the energy load of transformer $i$ on day $d$ as defined in
Sections \ref{sec:model} and \ref{sec:apply}. We use the same $\phi$'s
as $\psi$'s, a set of nine cubic B-spline basis functions with equally
spaced knots.  We consider the case where $\Sigma_c$ = \sigmagc${\I
 }$, $c=1, 2, 3$, and $\sigma_i^2 = \sigma^2$, $i=1,\ldots, I$.

We find initial estimates of $\sigma^2$, \sigmagum,
 \sigmagdois and \sigmagtres
using the methods described
in Section \ref{sec:apply}.  Specifically,
we calculate the ${\bf\hat{Y}}_{id}$'s using cubic smoothing splines.  We choose one  smoothing parameter by eye, to use for all spline fits.
The chosen smoothing parameter
 results in using 10 degrees of freedom to fit each replicate.    We find $\hat{\sigma}^{2(0)}$
via
\begin{equation}
\hat{\sigma}^{2(0)} = \frac{ \displaystyle\sum_{i=1}^{3}\sum_{d=1}^{5} \sum_{j=1}^{96}  \| \vecY_{id}[j]
- {\hat{\vecY}_{id}[j] } \| ^2  }{15(96 - 10)} = 7.16
.
 \label{eq:sigmag_data}
\end{equation}
To find our initial estimates, we suppose that \sigmagdois =  \sigmagum and \sigmagtres = $\sigma_{\gamma,1}^2$, which yields
 $\hat{\sigma}^{2(0)}_{\gamma,1} = \hat{\sigma}^{2(0)}_{\gamma,2} = 0.074 $ and
$\hat{\sigma}^{2(0)}_{\gamma,3} = 0.446$.
After our iterative maximization of the likelihood, we obtain our final estimates for the variance parameters:  ${\hat\sigma^2} = 7.17$,
$\hat\sigma^2_{\gamma,1} = 0.102$, $\hat\sigma^2_{\gamma,2} =
0.011$ and $\hat\sigma^2_{\gamma,3} = 0.338$.

Table \ref{tab:dataset} presents the number of reported consumers from
the residential and commercial classes and our estimates of the true
numbers of consumers.  According to our estimates, transformers 1 and
2 have the correct number of reported consumers for all classes, while
transformer 3 has one monophasic consumer reporting as biphasic.

The estimated typical energy usage curves of residential and
commercial consumers are shown in Figure \ref{fig:data_estimates}.
Panel (a) shows the three estimates together while panels (b), (c) and
(d) show each estimate separately.  We can see that,
at all times, the residential biphasic usage is higher than the
monophasic residential usage. Commercial usage drops close to zero
around 5am.  The peak load of commercial consumers is almost six times
the peak load of monophasic residential consumers and four times that of
biphasic residential consumers.

Residential consumers have a high peak of energy consumption around 8--9 pm, due to the
local habit of taking showers at night. There is another peak around
noon: if it is ``real", it probably occurs because Brazilians return
home for lunch.  Commercial consumers have a load that increases
between 5 am until a peak at around 6 pm.

Note that we do not have standard error bars for these estimates, and our statements cannot be made with quantifiable certainty.

To check how well our procedure fits the data,  we estimated the electric load through
the weighted sum of typical curves $\big(\sum_{c=1}^3 \hat{M}_{ci}
\hat{\alpha}_c(t)\big)$.  In Figure \ref{fig:aggregated_curves}, we plot this estimated function along with the observed electrical load.   We see
that we have obtained a very good fit.

\section{Simulation studies}
\label{sec:simulation}
We carry out eight different simulation studies, generating 200 data sets in each study.   For each  data set, we generate data from five transformers, each serving 75 consumers of two classes: residential ($c=1$) and commercial ($c=2$).    Four of the eight simulation studies contain replicates of power consumption curves for each transformer and four do not.  We consider two scenarios for typical consumer usage curves: one with the curves similar to what is expected in real data (with $\alpha_2$ much larger than $\alpha_1$) and one with the two curves similar in scale.  We also consider two scenarios for the
values of the $M$'s:  one similar to the data set, with the $M_1$'s much larger than $M_2$'s in all transformers (unbalanced $M$'s)  and one with the
sum of the $M_1$'s  across the five transformers approximately equal to the sum of the  $M_2$'s (balanced $M$'s).

Details of the scenarios and data generation are given in Section \ref{sec:data.generation}.    Results of the simulation studies are given in Section \ref{sec:simulation_results}.

\subsection{Data generation}
\label{sec:data.generation}
 In each of our eight simulation studies, energy consumption for each transformer is ``observed" every 15 minutes, so that there are 96 measurements taken each day, with time $t \in[0,24]$.
All data sets are generated from
equations (\ref{eq:alpha}), (\ref{eq:alphastar})  and  (\ref{eq:Yit}) with $\mbox{Var}(\epsilon_i(t)) = 3.5$,
  for $i=1,...,5$.
 For the basis functions, we use the same $\phi$'s as $\psi$'s, a set of nine cubic B-splines with equally spaced knots.
For consumer $l$ of class $c$ served by transformer $i$,  we construct  $\alpha^*_{cli }$ by sampling $\bgamma^{cli}$ from a multivariate normal distribution with mean of zero and
covariance matrix $\Sigma_c = \sigma^2_{\gamma,c}{\I}$.
 We  choose \sigmagc and $\alpha_c$, the class $c$'s expected energy consumption,  according to two different cases, one with ($\alpha_1, \sigma_{\gamma,1})$ the
 same scale as ($\alpha_2, \sigma_{\gamma,2}$) and the other with the scale of ($\alpha_1, \sigma_{\gamma,1}$) much smaller than the
scale of ($\alpha_2, \sigma_{\gamma,2}$). Details are given below.

For the $i$th transformer, we generate the reported counts in the two classes by generating multinomial vectors $\vecX_{i1}$  and $\vecX_{i2}$ as
described in Section \ref{sec:model.Ri}, using the fraud matrix given in (\ref{eq:Fraud}).    For each transformer, we
choose to generate the reported counts once and use these reported counts  in all simulation studies.
Table \ref{table:true.counts} contains the true and reported counts for consumers of class 1 for both types of $M$'s -- balanced $M$'s and unbalanced $M$'s.
We use the same reported counts in all simulated data sets because it is easier to study the properties of the estimated true counts, as we can see in Table  \ref{tab:traf2-bal} of the paper and
Tables 3 and 4 of the supplementary material.

 We study  the four cases listed below.

\vspace{0.3cm}
{\bf Case 1:} The two functions $\alpha_1$ and  $\alpha_2$ are of the same scale and the $M_{1}$'s and $M_{2}$'s are balanced.

\vspace{0.3cm}
{\bf Case 2:} The two functions  $\alpha_1$ and $\alpha_2$ are of the same scale and the $M_{1}$'s are much bigger than the $M_{2}$'s.

\vspace{0.3cm}
{\bf Case 3:} The function  $\alpha_1$ is of a much smaller scale than the function  $\alpha_2$ and the $M_{1}$'s and $M_{2}$'s are balanced.

\vspace{0.3cm}
{\bf Case 4:} The function  $\alpha_1$ is of a much smaller scale than the function  $\alpha_2$ and the $M_{1}$'s are much bigger than the $M_{2}$'s.

\vspace{0.3cm}

We do not consider the case where $\alpha_1$ is of a much smaller scale than  $\alpha_2$ and the $M_{1}$'s are much smaller than the $M_{2}$'s, as the estimates of
$\alpha_1$ and $\alpha_2$ are extremely poor in this challenging case.

To relate these cases to our example, consider class 1 as residential
and class 2 as commercial.  We would expect the load of a consumer of
residential class to be lower than that of a consumer of commercial
class, that is, $\alpha_1$ is of smaller scale than $\alpha_2$, as in
Cases 3 and 4.  Also, typically, the number of residential consumers
served by a transformer is higher than the number of commercial
consumers (Cases 2 and 4).

Figures \ref{fig:quant-case1} and \ref{fig:quant-case2} show the curves $\alpha_1$ and $\alpha_2$ for Cases 1 and 2. Figures \ref{fig:quant-case3} and \ref{fig:quant-case4} present the curves for Cases 3 and 4.

For Cases 3 and 4 we obtain $\alpha_1$ and $\alpha_2$ along with the corresponding $\bgamma^1$ and $\bgamma^2$ by fitting a B-spline model to the data considering a residential class $c=1$ and a commercial class $c=2$. For Cases 1 and 2,
we rescale $\alpha_1$ and $\alpha_2$ by rescaling the associated
$\bgamma^1$ and $\bgamma^2$ so that all components are between 0 and 1
and the simulated curves are all positive. For instance, to rescale
$\alpha_1$, let $a_1$ equal the minimum of $\bgamma^1$'s components
and $b_1$ equal the maximum. We define the rescaled $\alpha_1(t)$ as
$\bphi(t)' \bgamma^{*1} + 2$ with
\begin{equation}
\gamma^{*1}[k] = \frac{\hat{\gamma}^{1}[k] -  a_1}{b_1-a_1} .
\end{equation}

For  the variance parameters for the consumer level energy consumption curves, for Cases 1 and 2, we set \sigmagum = 0.03 and \sigmagdois = 2 $\times$ \sigmagum = 0.06.
For Cases 3 and 4,  to maintain the relative  variability of the consumer level curves about the $\alpha_c$'s, we rescale \sigmagc  by multiplying  by  the appropriate constant: $(b_c-a_c)^2$,
$c=1,2$.
In Cases 1--4, the value of $ \sigma_{\gamma,2}^2$ is twice  $\sigma_{\gamma,1}^2$ because we believe that this reflects the  variability
among commercial consumers compared to the variability among  residential consumers.

Figure \ref{fig:simulation_data1} shows an example of simulated  consumer-level data for Case 1 for the first transformer. The left plot shows the energy consumption of 5 out of the 45 consumers of class 1 (residential consumers) and the right plot shows the energy consumption of 5 out of the  30 consumers of class 2 (commercial consumers). Recall that in practice these consumer level consumption curves are not observed:  we only observe the sum of all of the 75 curves.

For each of Cases 1--4, we generate two types of data sets: one with
just one day observed per transformer and another with five days (replicates)
observed per transformer.  We generate the days independently, which
is a simplification since consumer level day to day usages are
probably correlated.

\subsection{Analysis and Results}
 \label{sec:simulation_results}
In calculating our estimates, we consider the same fraud matrix $F$ and the same basis functions used to generate data. We also assume
that  $\Sigma_c = \sigma^2_{\gamma,c}{\bf I}$ and $\sigma_i^2 = \sigma^2$.
 Recall that for the maximization in Step 2 we require values of the $H$ function given in (\ref{eq:Hfunction}). As described in Section  \ref{sec:R.given.M},
 we approximate $H$ in each transformer by simulating $B=100,000$ independent data sets just once and tabling the results.

The estimates of $\alpha_1$ and $\alpha_2$ are summarized in Figures \ref{fig:quant-case1}-\ref{fig:quant-case4}, which  show the pointwise minimum, maximum, median and quartiles
of the 200 estimates for each of Cases 1--4.
We see that estimates of the  $\alpha_c$'s  are best  when the true $\alpha_c$'s are of the same scale (Figures  \ref{fig:quant-case1} and \ref{fig:quant-case2}).  As expected, estimates of the $\alpha_c$'s are less variable when they are based on data with replicates (bottom row of plots).   When $\alpha_1$ is much smaller than $\alpha_2$, the estimates are severely biased
(Figures \ref{fig:quant-case3} and \ref{fig:quant-case4}).
We always obtain excellent estimates of the total electric load for each transformer through the weighted sum $\hat{M}_{1i} \hat{\alpha}_1(t)
+ \hat{M}_{2i} \hat{\alpha}_2(t)$, $i=1,\ldots, 5$ (e.g., see Figure \ref{fig:agregCase4_example}).

Estimates of true consumer counts in the five transformers are
provided in Tables 3 and 4 of the supplementary material.  The results for transformer $i=2$ are reproduced here in Table \ref{tab:traf2-bal}.
 The estimates of
$M_1$ are tabled in the top half of Table \ref{tab:traf2-bal} for
cases with $\alpha_1$ and $\alpha_2$ of the same scale (Case 1) and of
different scales (Case 3), for data with or without replicates.
For instance, in the balanced $M$ case (Case 1),
transformer 2 contains $M_1=$ 29 consumers of class $c=1$, with 32
consumers reporting that they are of class $c=1$.
We
see that in all cases in this transformer, the most common estimate
is $\hat{M}_1=31$.  We also see that the estimates of $M_1$ are more
variable for data with no replicates, as expected.

In looking at Tables 3 and 4 of the supplementary material, we see that in all cases, the most common estimate of $M_1$ is either at the reported number or at the true value of $M_1$ or at some number in between.  In many cases, the most common value is equal to the truth, or at least shifted from the reported number towards the truth.   Notable exceptions are in transformer 3 in Table 3, when the $M$'s are balanced and in transformers 3, 4 and 5 in Table 4 when the $M$'s are unbalanced. In these exceptions,
the most common value of the estimates is the reported number (with one
exception).  An extreme case of this is in Table 4 (the unbalanced $M$'s case):   in transformer 4, our
estimates of $M_1$ seem ``stuck" on the value of the reported number of class
1 consumers. This bias towards the reported number of consumers of class 1 may be due to the fact that, in some cases, the reported number  is lower than the true number, which is
unusual.  As expected, the variability of estimates is always lower when the data contain replicates but, perhaps surprisingly, the bias is not lower.

 For variance component estimation, the regression error variance is
 very well-estimated in all simulation scenarios.  However, estimates of the variance parameters
 of the consumer level energy consumption curves caused some problems,
 particularly when we observe only one day of data  per transformer.  In fact,
 in this case, in many data sets, the estimated consumer level
 variances are equal to zero.

To investigate the effect of increasing the number of transformers as well as
increasing the number of replicates (number of days)  we conduct further simulations and present the results in 
the Supplementary Material.  We consider the following scenarios: (i) 5 transformers and 30 days of
observation; (ii) 5 transformers and 100 days of observation; (iii)
50 transformers and 1 day of observation; (iv) 50 transformers with 5
days of observation. As expected, the variability of the estimated typologies is reduced by increasing the number
of transformers and/or the number of replicates.  In addition, the bias  in  typology estimation is reduced by increasing the number of transformers.
Perhaps surprisingly, this bias is not reduced by increasing the number of replicates.  Thus, in cases where estimates have a large bias and a
decreased variability, the estimated typologies concentrate around the wrong curve.
Even with the increase number of replicates and/or number of transformers, in Cases 3 and 4 (when $\alpha_1$ is much smaller than $\alpha_2$)
many estimates of $\sigma_{\gamma,1}^2$  were zero.  Estimation of $\sigma^2$ was good throughout. In general, we found that estimation of $M_1$ was as described above:  the estimates tended to be shifted from the reported number of consumers of class 1 to the true number.
The variability of the estimates decreased with the number of replicates, but surprisingly, not with the number of transformers.  The bias typically did not decrease when we increased  the number of replicates but did usually decrease with an increase in the number of transformers.

%%%%%%%%%%%%%%%%%%%%%%%%%%%%%%%%%%%%%%%%%%%%%%%%%%%%%

\section{Conclusions}

In this paper we proposed a generalization of the work of
\cite{dias:garcia:martarelli:2009} on estimating mean curves
when the available sample consists of aggregated functional data.
The main novelty of this work is to
incorporate a randomness in the counts for class membership. This
flexibility allows the analysis of the data even when there is some
misreporting in the number of consumers of each type. We also use random effects to model the within transformer correlation structure.

To study the properties of our method, we analyzed artificial data sets
exploring different scenarios and we also analyzed a real data set. The artificial data
allowed us to explore the influence of increasing the number of
replications and the number of transformers. In the data example, it is clear from comparing the observed
aggregated curves with the weighted sum of the estimated typical ones,
that the proposed model  provides reasonable estimates of the  mean curve.

\section*{Supplementary material}

The file \texttt{supplementary.pdf} contains supplementary plots,
tables and comparisons for the examples discussed in this paper. It
also includes the results of further simulation studies for different
numbers of transformers and replicates.

%%%%%%%%%%%%%%%%%%%%%%%%%%%%%%%%%%%%%%%%%%%%%%%%%%%%%%

\bibliographystyle{imsart-nameyear}
\bibliography{referbib}

\clearpage
%%\vfill\eject
%%%%%%%%%%%%%%%%%%%%%%%%%%%%%%%%%%%%%%%%%%%%%%%%%%%%%
\begin{figure}
\centering
\includegraphics[width=10cm]{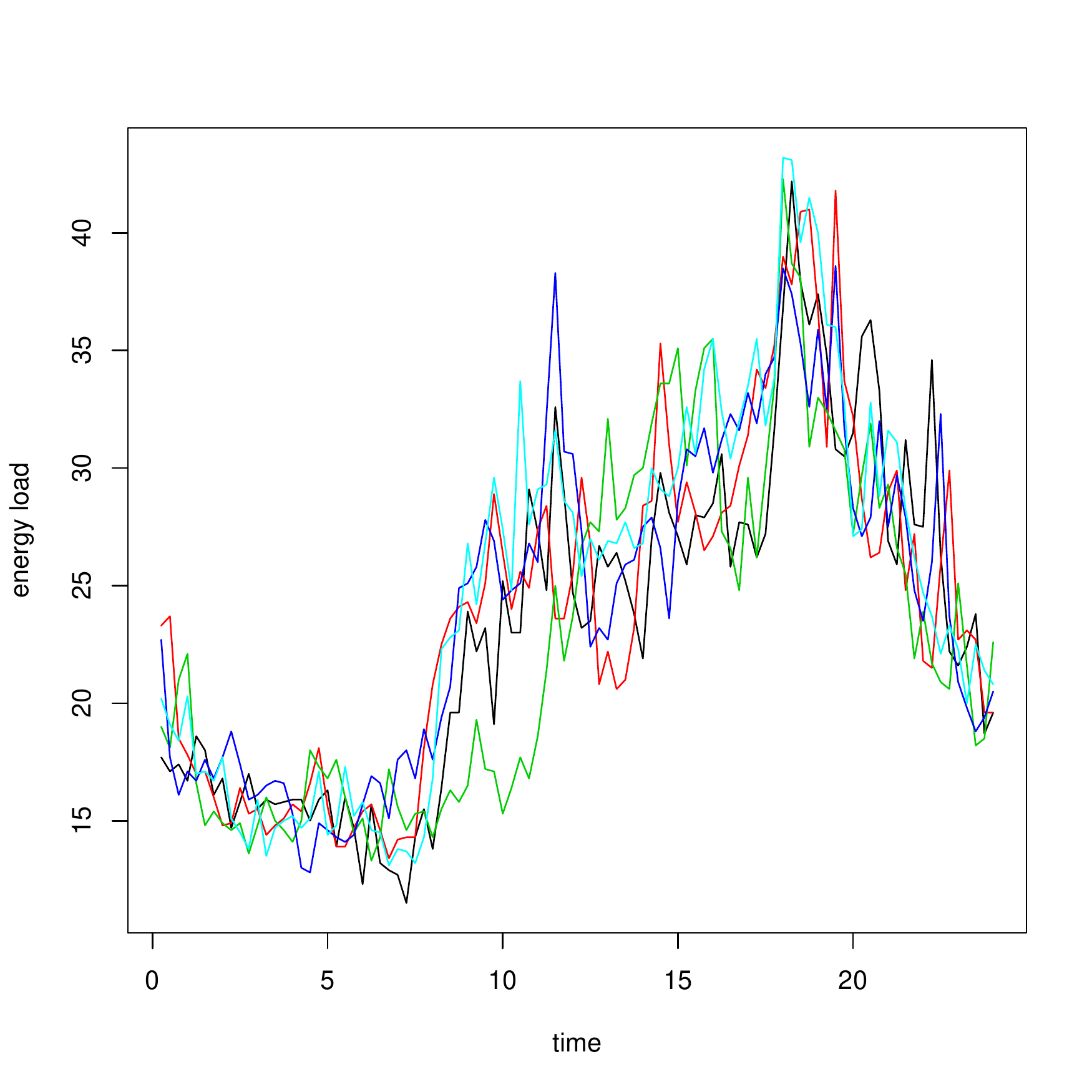}
\caption{Data analysis: data from Transformer 1. The plot shows total electricity usage for each of five weekdays between 06/21/2002 and 06/27/2002. Each color corresponds to a different day.}
\label{fig:data.example}
\end{figure}

\clearpage
%%%%%%%%%%%%%%%%%%%%%%%%%%%%%%%%%%%%%%%%%%%%%%%%%%%%%
%
\begin{figure}
        \centering

        \begin{subfigure}[b]{0.45\textwidth}
                \centering
         \includegraphics[width=\textwidth]{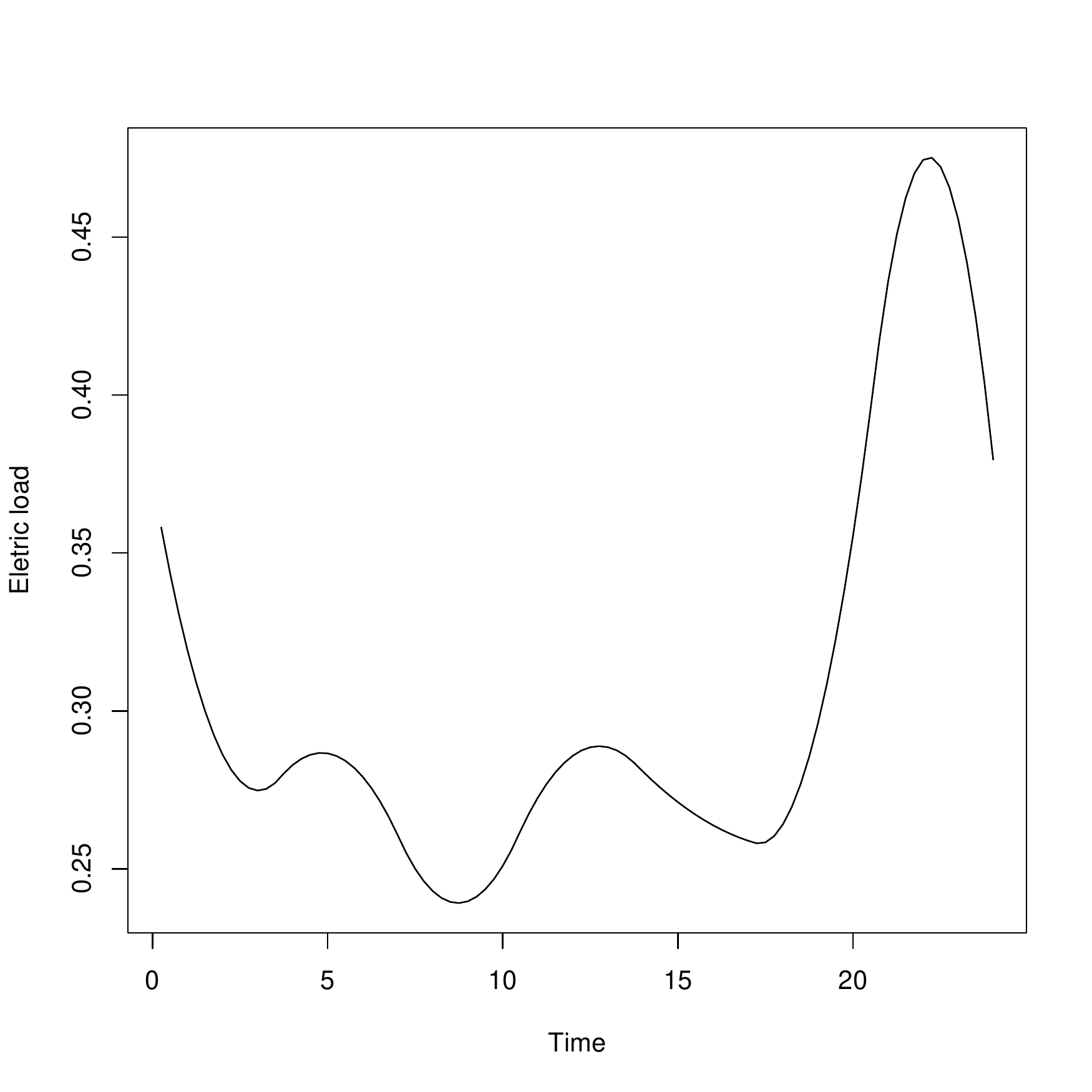}
                \caption{$\hat{\alpha}_1$}
                %%\label{}
        \end{subfigure}
        \begin{subfigure}[b]{0.45\textwidth}
                \centering
         \includegraphics[width=\textwidth]{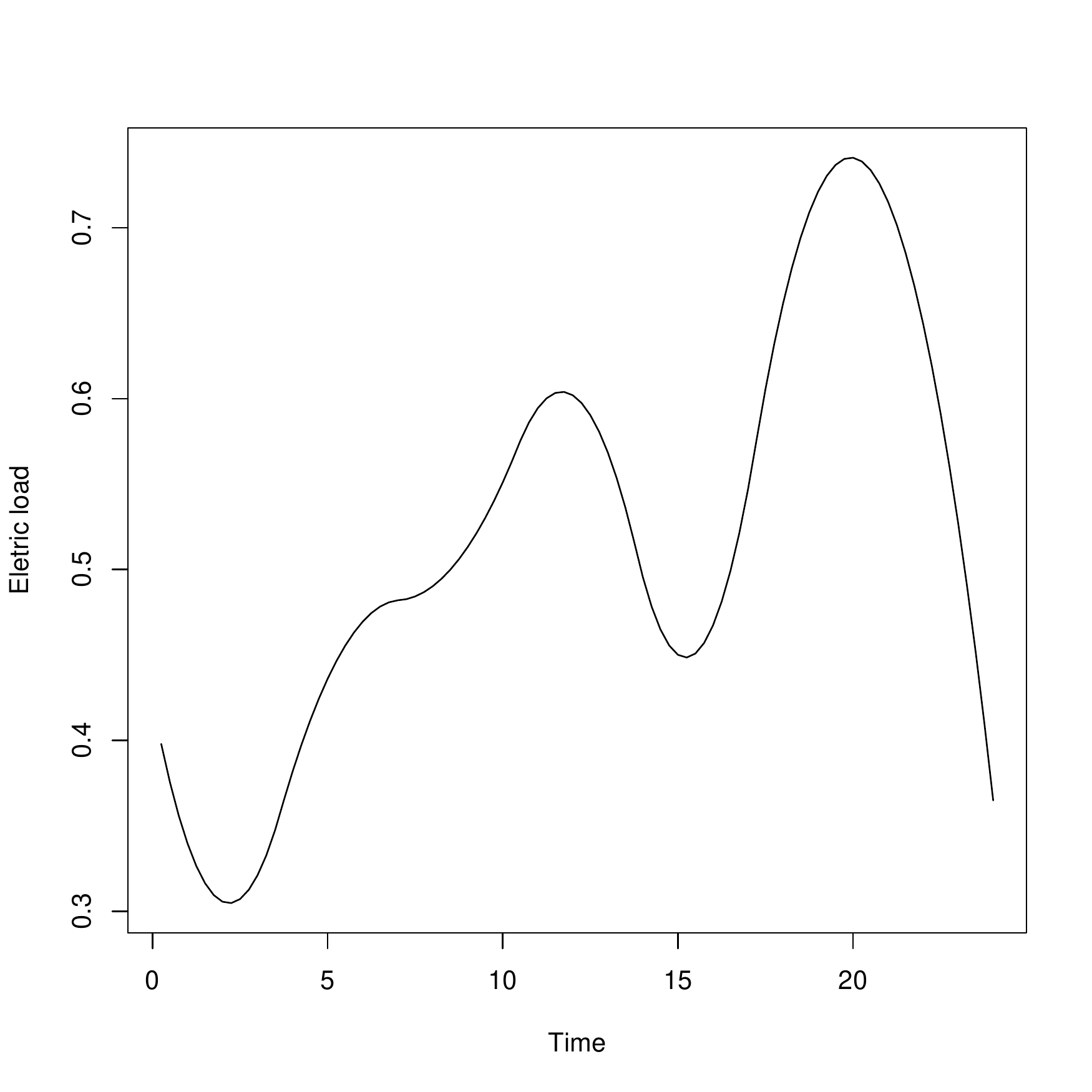}
         \caption{$\hat{\alpha}_2$}
        \end{subfigure}
        \vspace{.1cm} \\
        \begin{subfigure}[b]{0.45\textwidth}
                \centering
              \includegraphics[width=\textwidth]{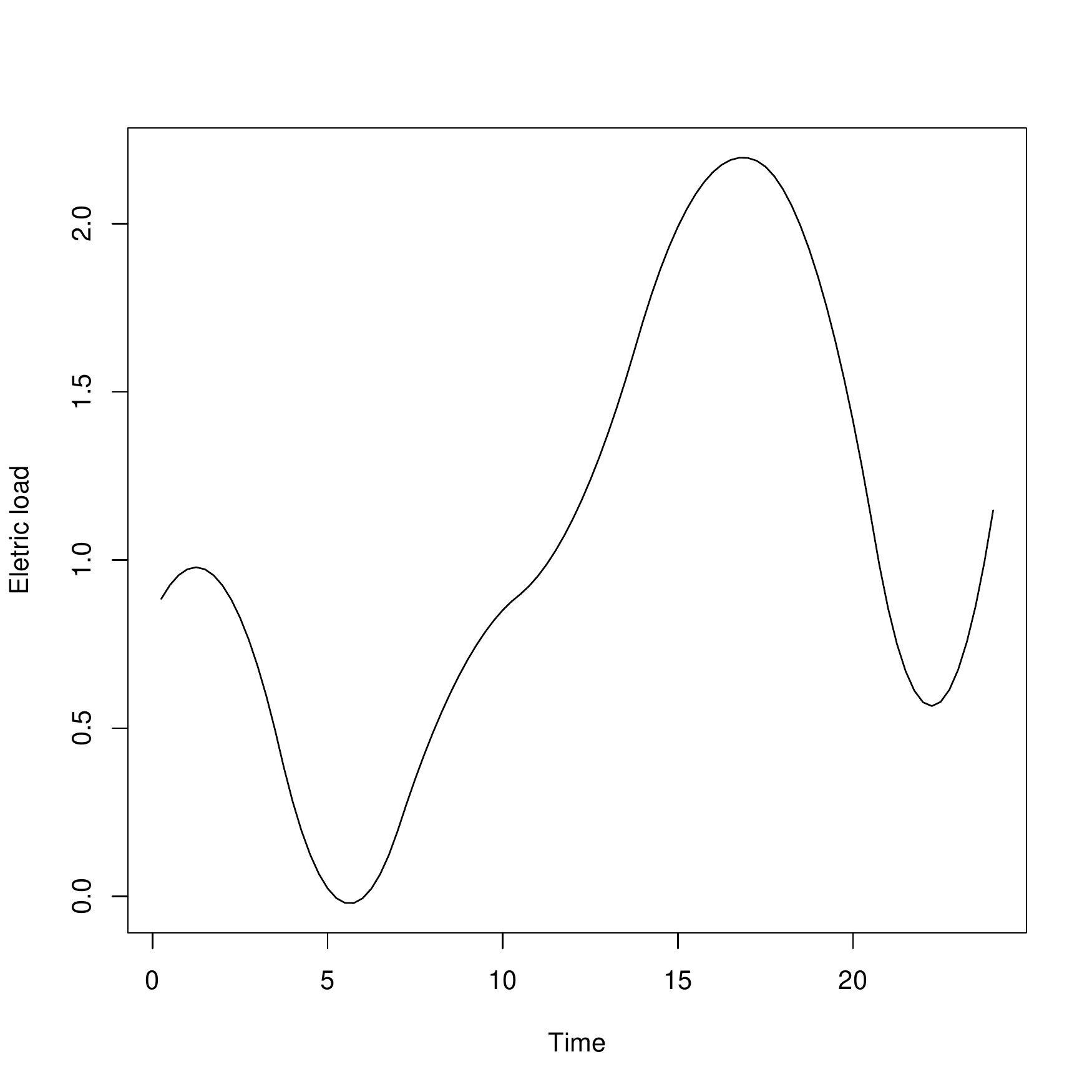}
              \caption{$\hat{\alpha}_3$}
                %%\label{}
        \end{subfigure}
        ~
        \begin{subfigure}[b]{0.45\textwidth}
                \centering
               \includegraphics[width=\textwidth]{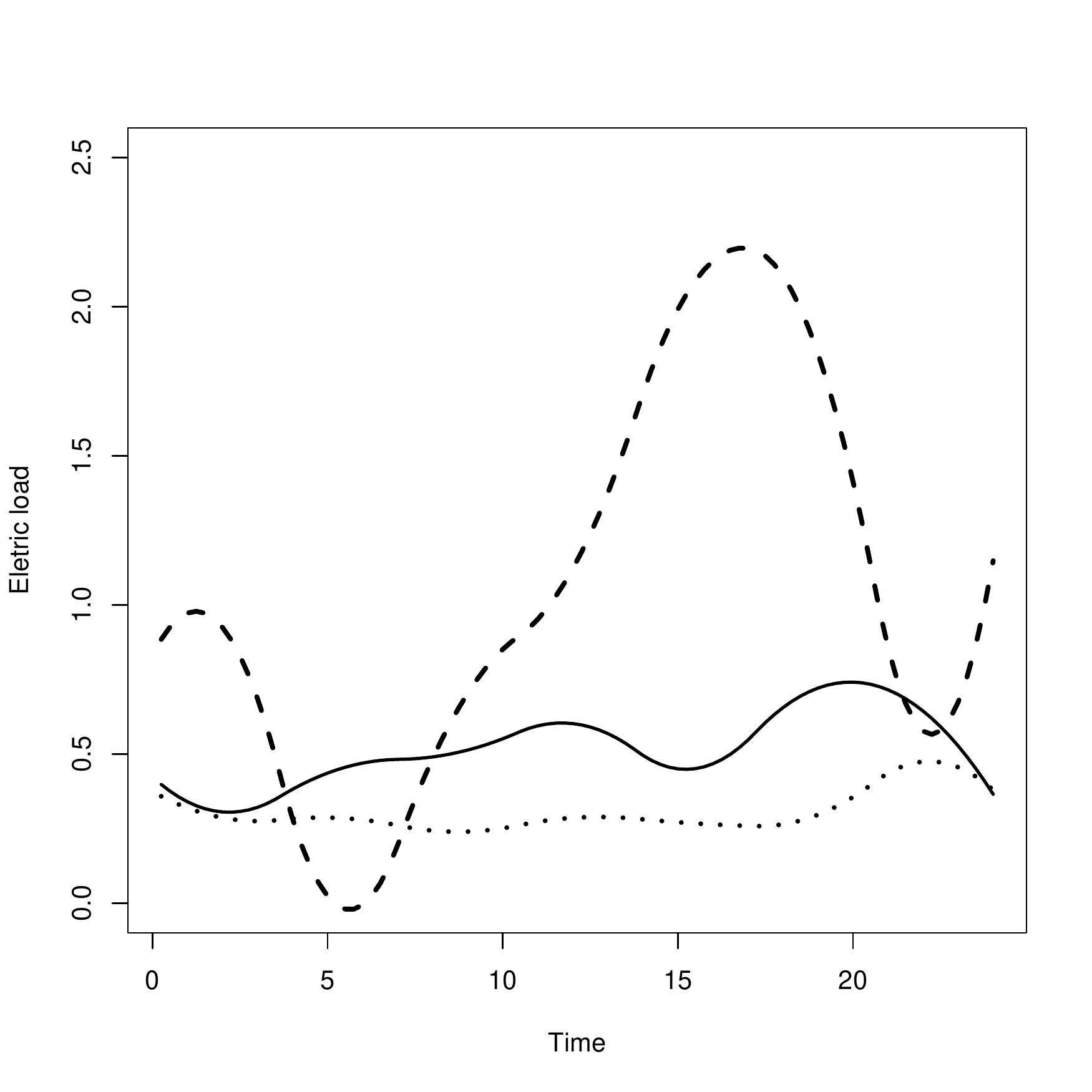}
             \caption{$\hat{\alpha}_1$, $\hat{\alpha}_2$ and $\hat{\alpha}_3$}
                %%\label{}
        \end{subfigure}
\caption{Data analysis: estimated expected energy consumption,
  $\hat{\alpha}_1$ (residential monophasic), $\hat{\alpha}_2$
  (residential biphasic) and $\hat{\alpha}_3$ (commercial). (a), (b) and (c) show $\hat{\alpha}_1$, $\hat{\alpha}_2$ and $\hat{\alpha}_3$, respectively. (d) shows the three estimated curves together, $\hat{\alpha}_1$ (dotted curve), $\hat{\alpha}_2$ (solid curve) and $\hat{\alpha}_3$ (dashed curve).}
\label{fig:data_estimates}
\end{figure}

%%~~~~~~ \vskip 20pt

\clearpage
%%%%%%%%%%%%%%%%%%%%%%%%%%%%%%%%%%%%%%%%%%%%%%%%%%%%%

\begin{figure}
        \centering

        \begin{subfigure}[b]{0.45\textwidth}
                \centering
         \includegraphics[width=\textwidth]{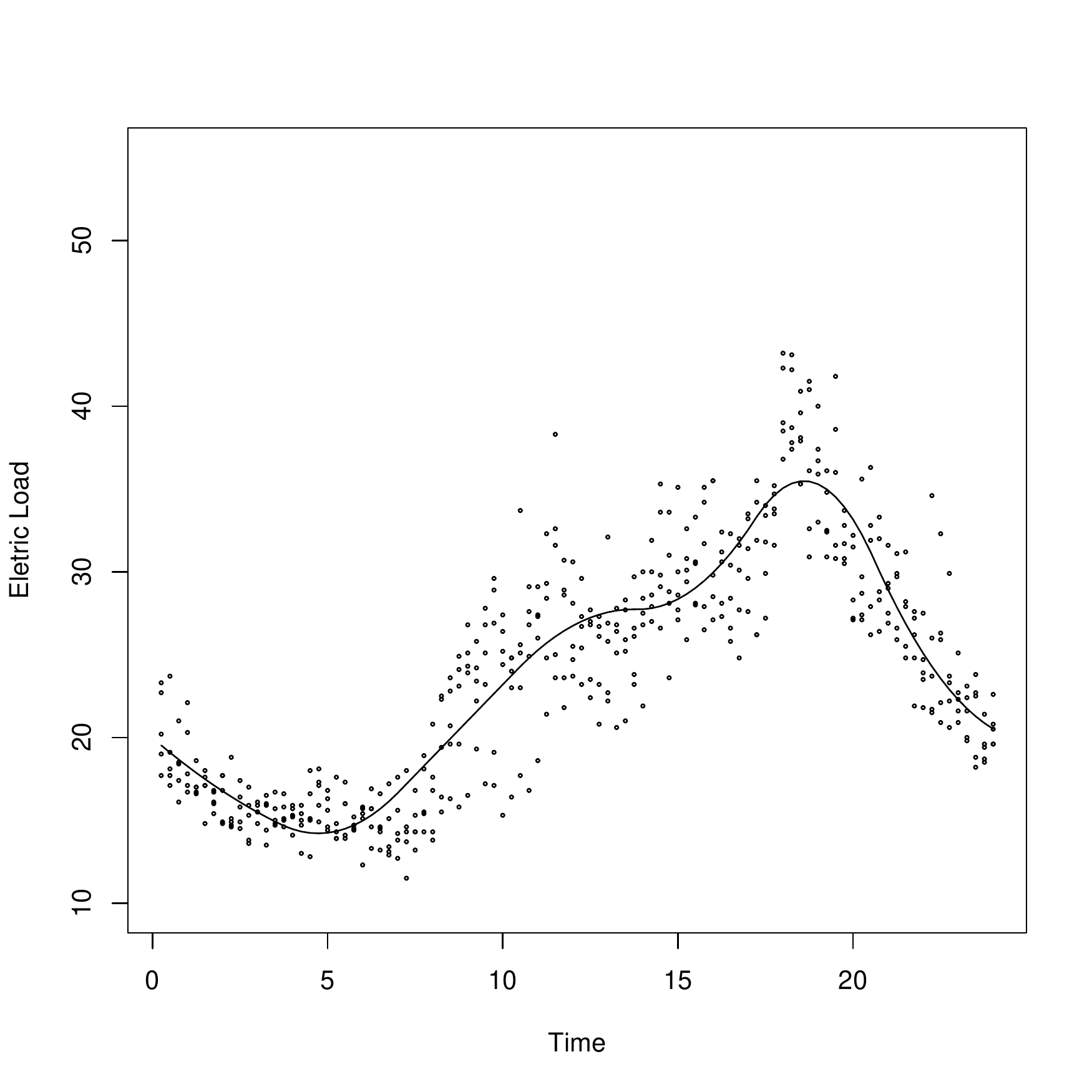}
                \caption{Transformer 1}
                %%\label{}
\end{subfigure}
        \begin{subfigure}[b]{0.45\textwidth}
                \centering
         \includegraphics[width=\textwidth]{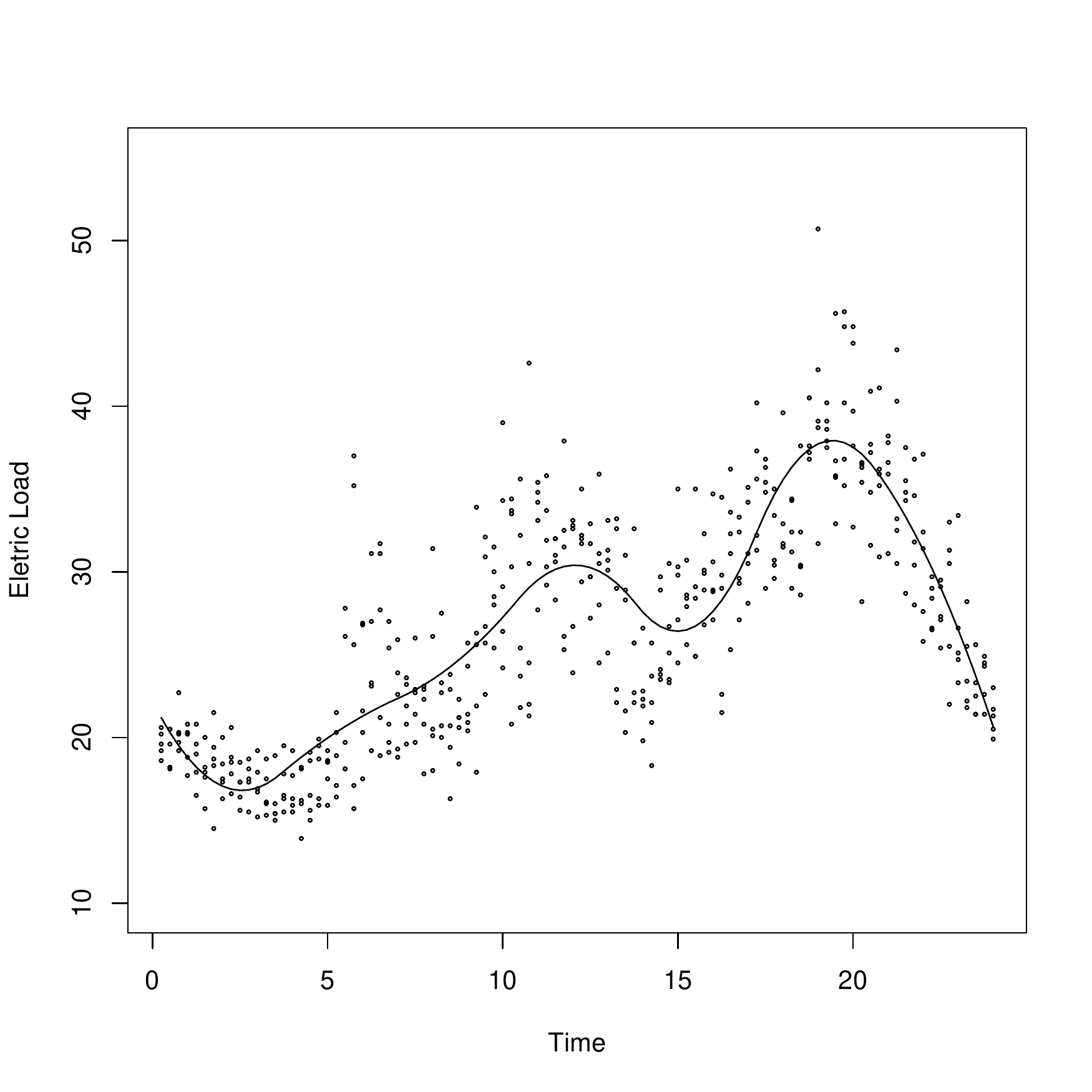}
         \caption{Transformer 2}
        \end{subfigure}
        \vspace{.1cm} \\
        \begin{subfigure}[b]{0.45\textwidth}
                \centering
               \includegraphics[width=\textwidth]{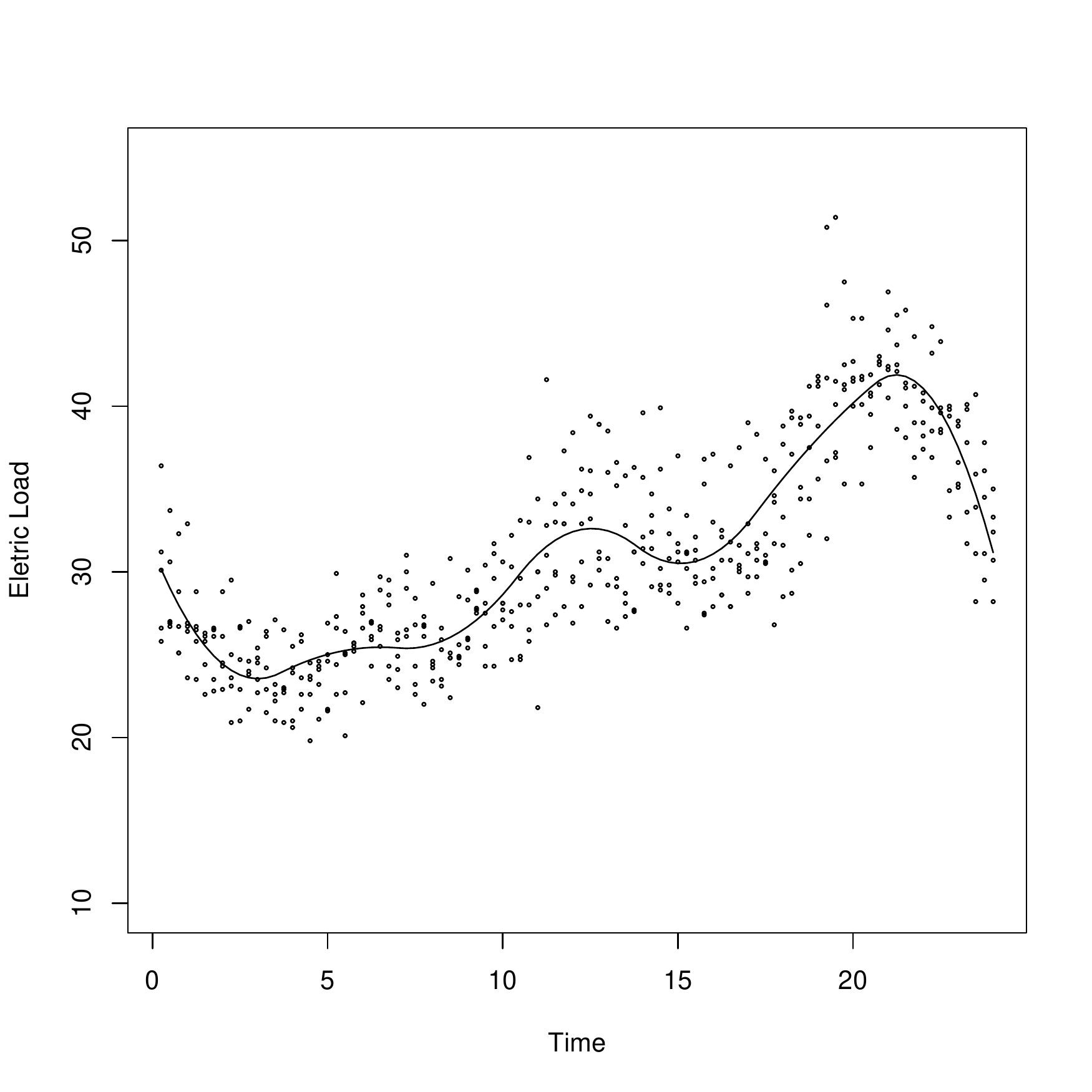}
               \caption{Transformer 3}
                %%\caption{}
                %%\label{}
        \end{subfigure}
        ~

\caption{Data analysis: observed data and estimated aggregated curves for transformers 1, 2 and 3.}
\label{fig:aggregated_curves}
\end{figure}

%%%%%%%%%%%%%%%%%%%%%%%%%%%%%%%%%%%%%%%%%%%

\begin{figure}[!htb]
\centering
\begin{subfigure}{0.5\textwidth}
\centering
  \includegraphics[width=\textwidth]{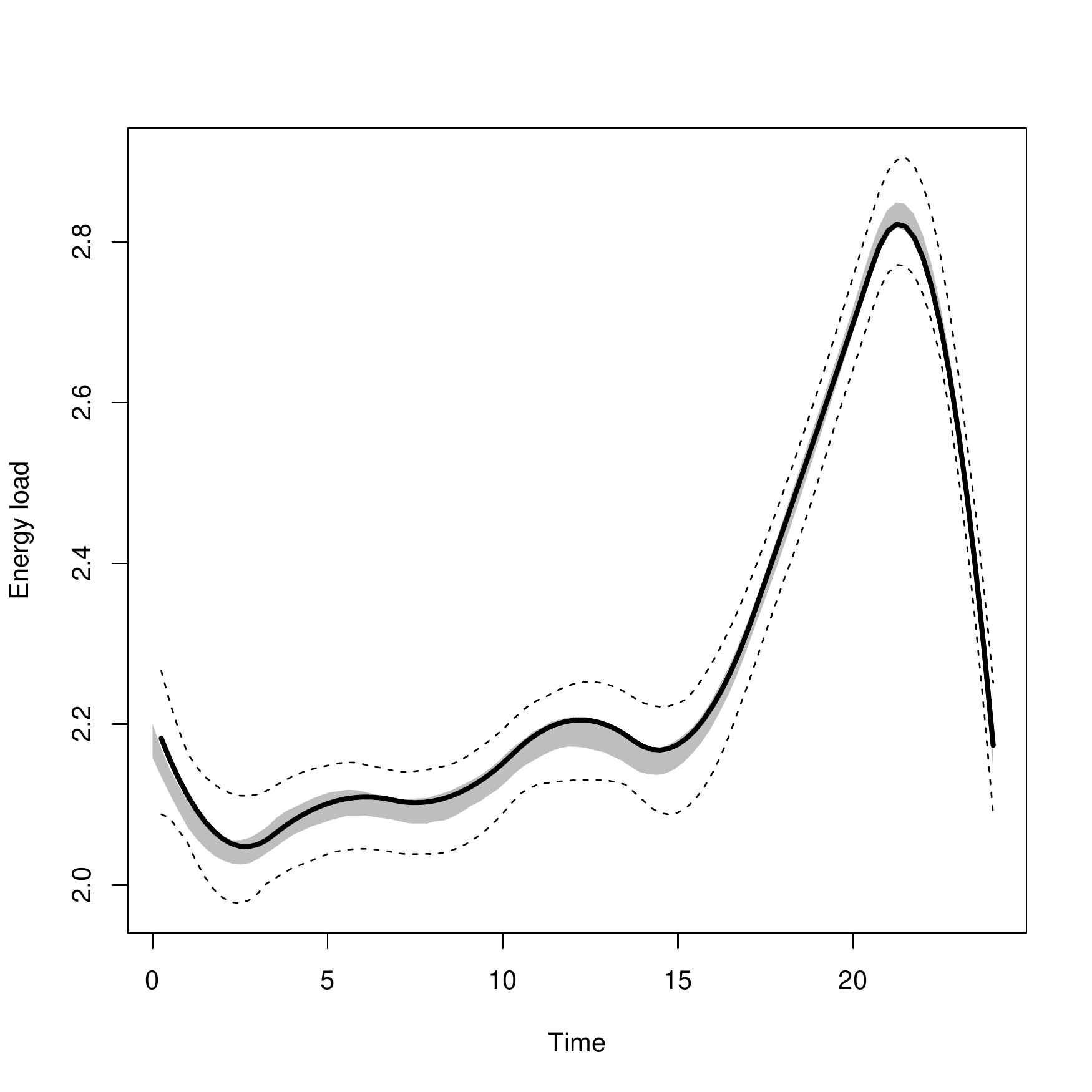}
\end{subfigure}%
\begin{subfigure}{0.5\textwidth}
\centering
   \includegraphics[width=\textwidth]{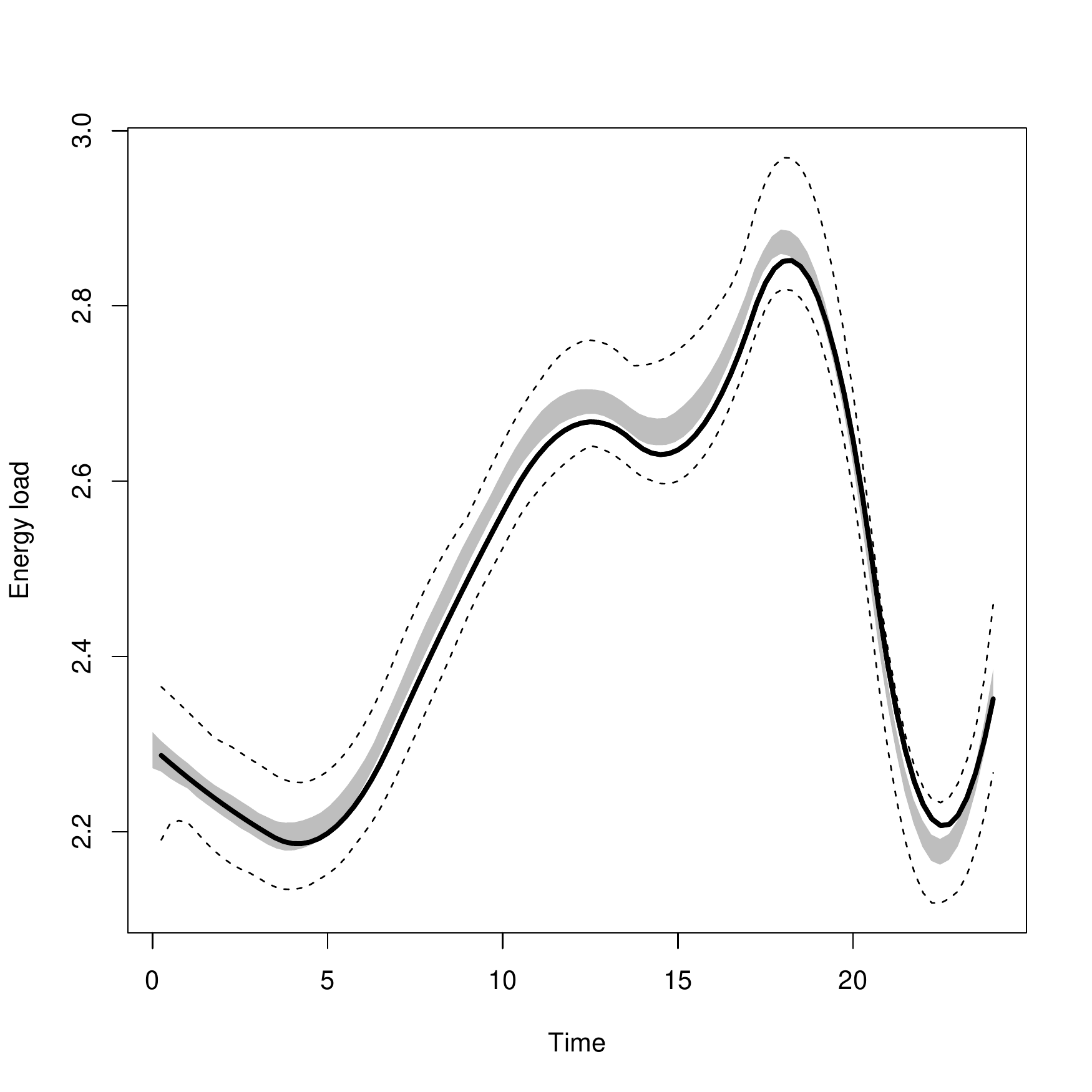}

\end{subfigure}
\begin{subfigure}{0.5\textwidth}
\centering
   \includegraphics[width=\textwidth]{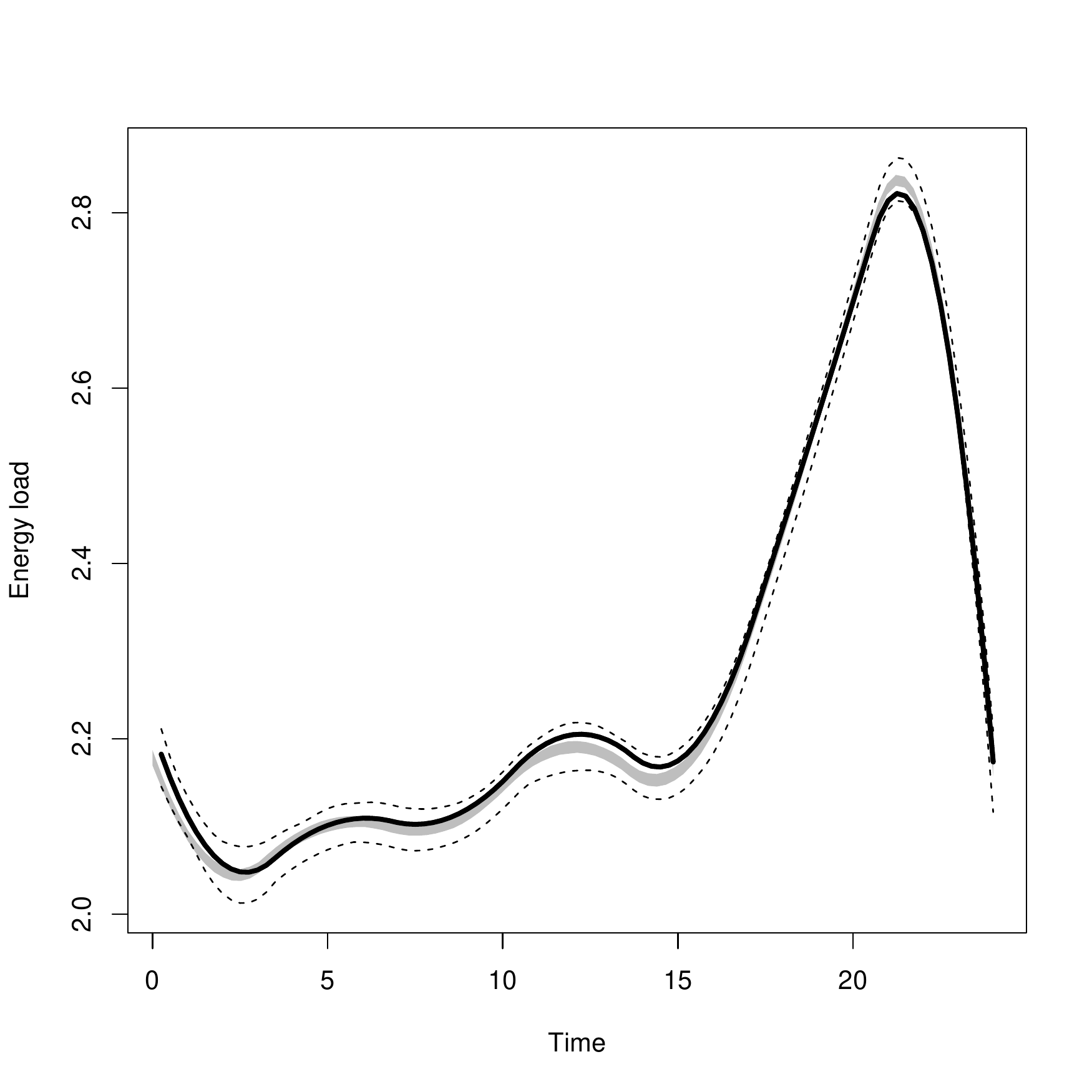}
\end{subfigure}%
\begin{subfigure}{0.5\textwidth}
\centering
   \includegraphics[width=\textwidth]{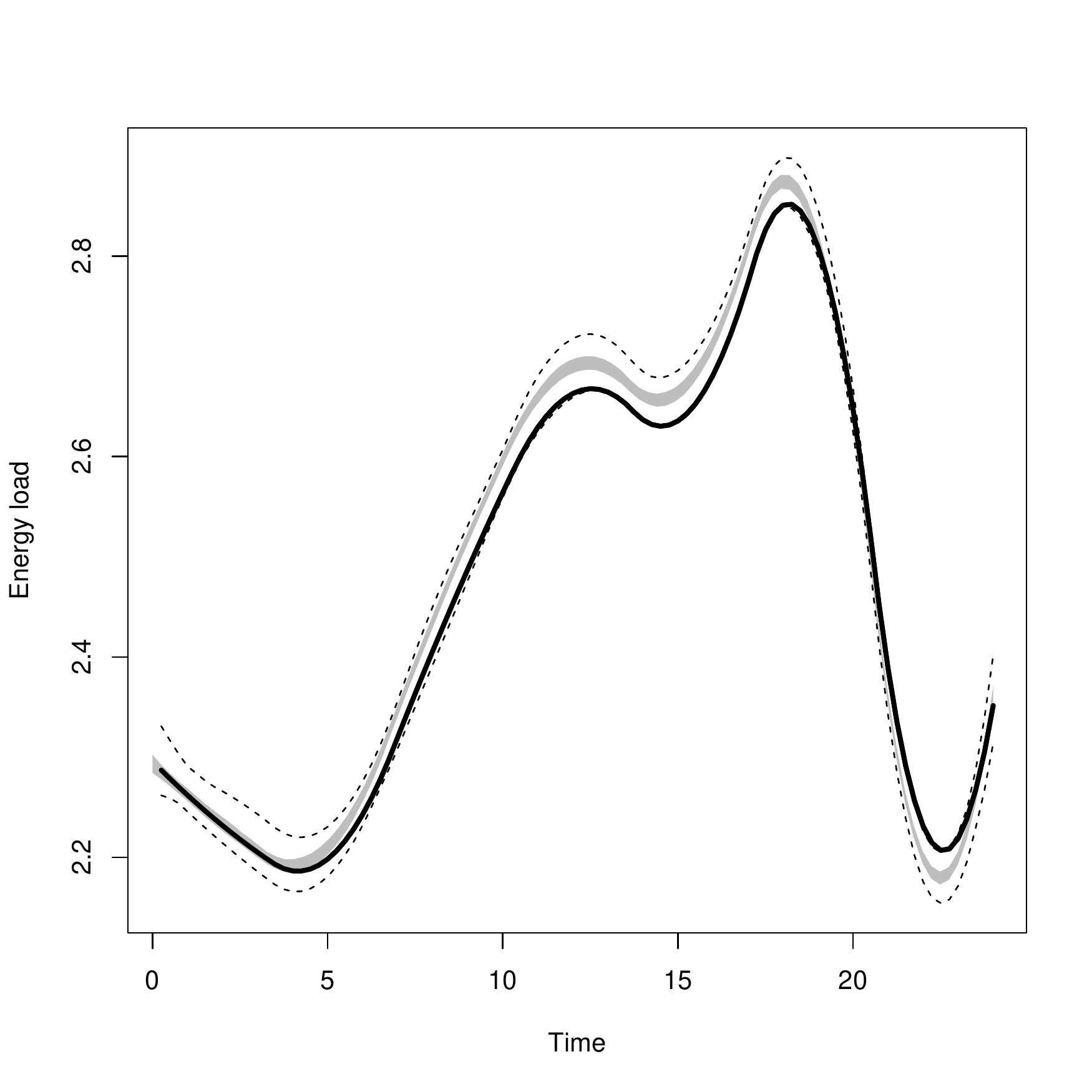}
\end{subfigure}
\caption{Simulation study Case 1 (\textit{$\alpha_1$ and $\alpha_2$ are of the same scale
and the $M$'s are balanced}): pointwise minimum, maximum, first and third quartiles
of the 200 estimated typologies for classes $c=1$ residential (left column) and $c=2$ commercial (right column) without replicates (top row) and with replicates (bottom row). The solid curve is the true typology used to generate the data, the shaded gray area corresponds to the area between the first and third quartiles and the dashed lines corresponds to the minimum and maximum.}

\label{fig:quant-case1}
\end{figure}

%%%%%%%%%%%%%%%%%%%%%%%%%%%%%%%%%%%%%%%%%%%

\begin{figure}[!htb]
\centering
\begin{subfigure}{0.5\textwidth}
\centering
  \includegraphics[width=\textwidth]{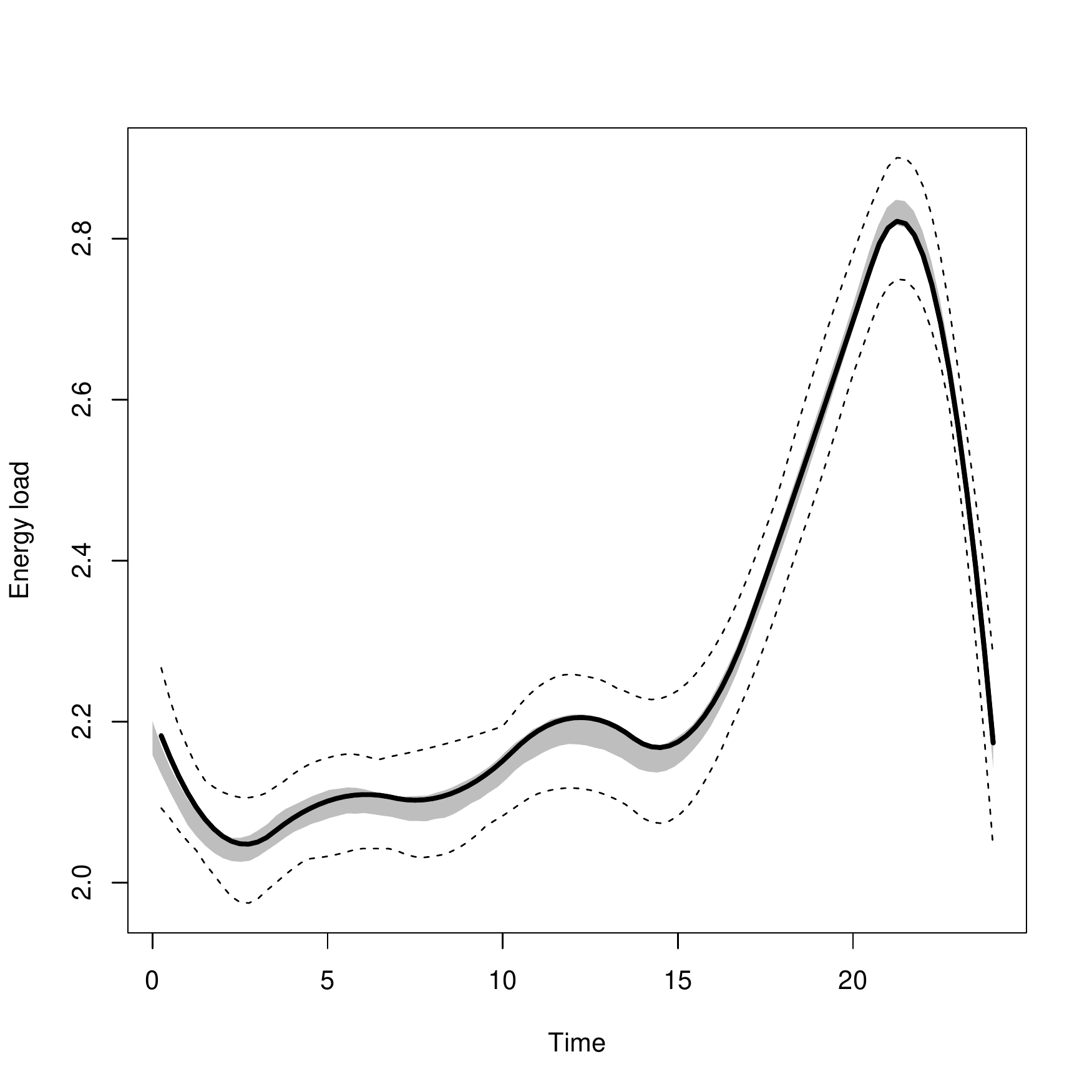}
\end{subfigure}%
\begin{subfigure}{0.5\textwidth}
\centering
  \includegraphics[width=\textwidth]{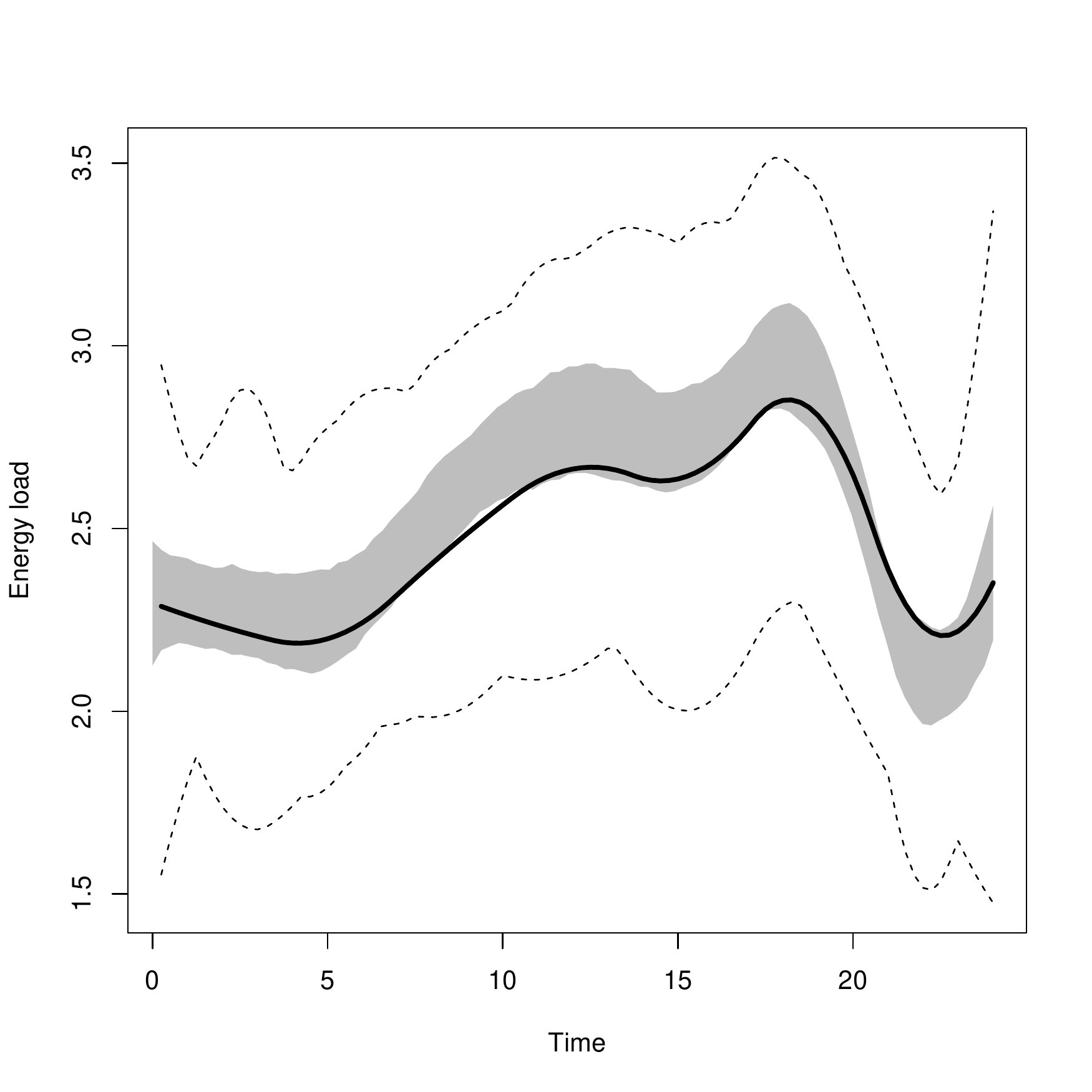}
\end{subfigure}
\centering
\begin{subfigure}{0.5\textwidth}
\centering
  \includegraphics[width=\textwidth]{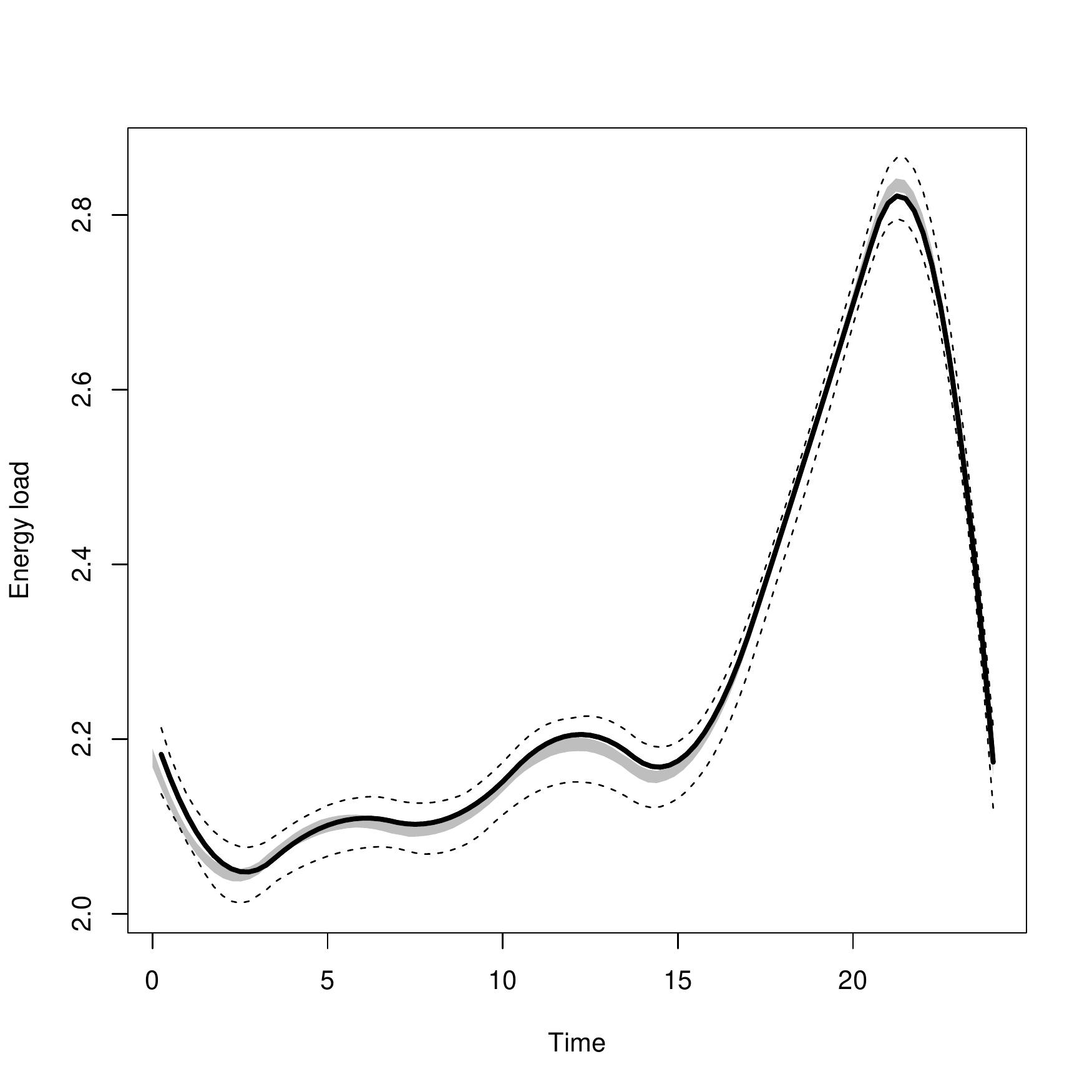}
\end{subfigure}%
\begin{subfigure}{0.5\textwidth}
\centering
  \includegraphics[width=\textwidth]{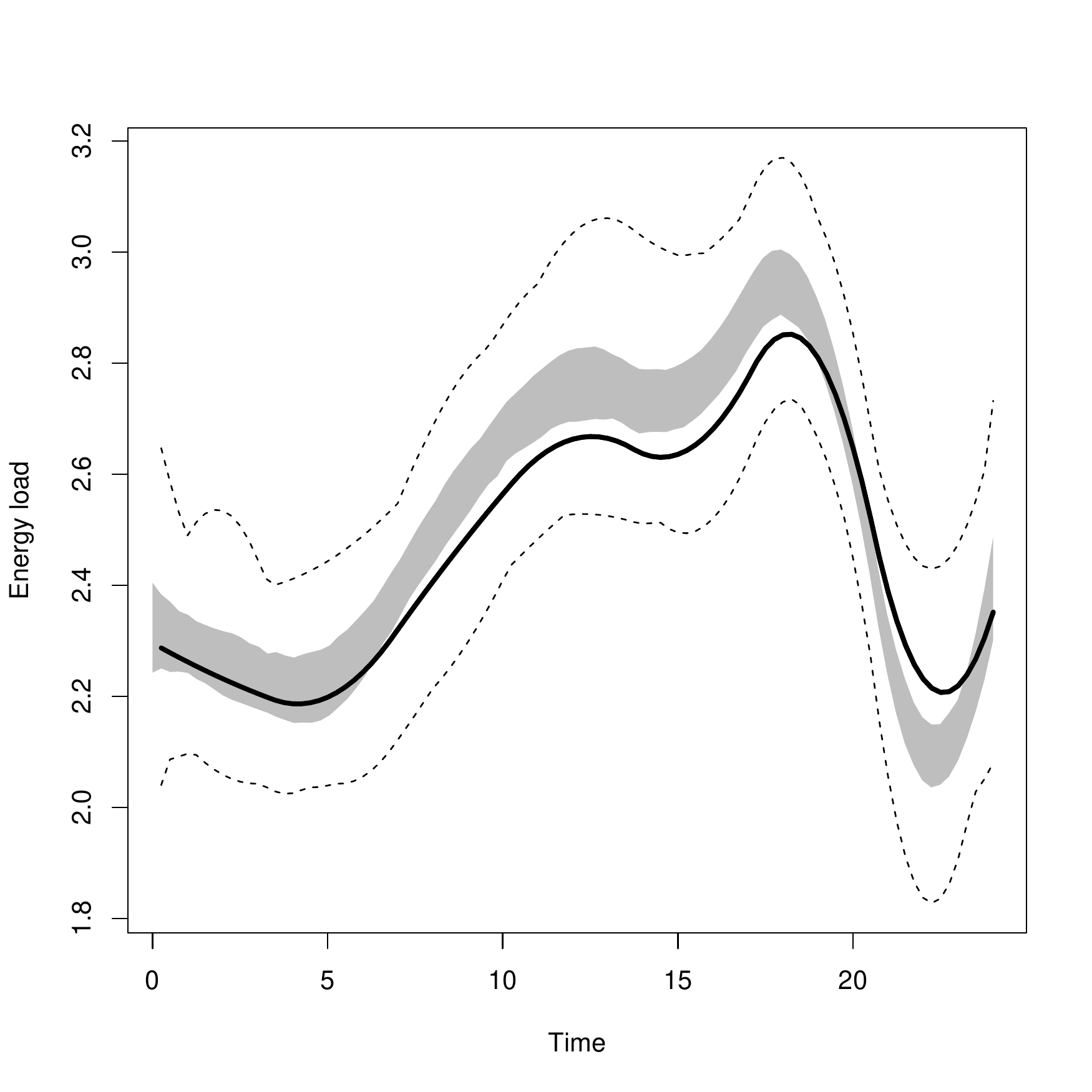}
\end{subfigure}
\caption{Simulation study Case 2 (\textit{$\alpha_1$ and $\alpha_2$ are of the same scale
and the $M_1$'s are much bigger than the $M_2$'s}): pointwise minimum, maximum, first and third quartiles
of the 200 estimated typologies for classes $c=1$ residential (left column) and $c=2$ commercial (right column) without replicates (top row) and with replicates (bottom row) as in Figure \ref{fig:quant-case1}.}
\label{fig:quant-case2}
\end{figure}

%%%%%%%%%%%%%%%%%%%%%%%%%%%%%%%%%%%%%%%%%%%

\begin{figure}[!htb]
\centering
\begin{subfigure}{0.5\textwidth}
\centering
  \includegraphics[width=\textwidth]{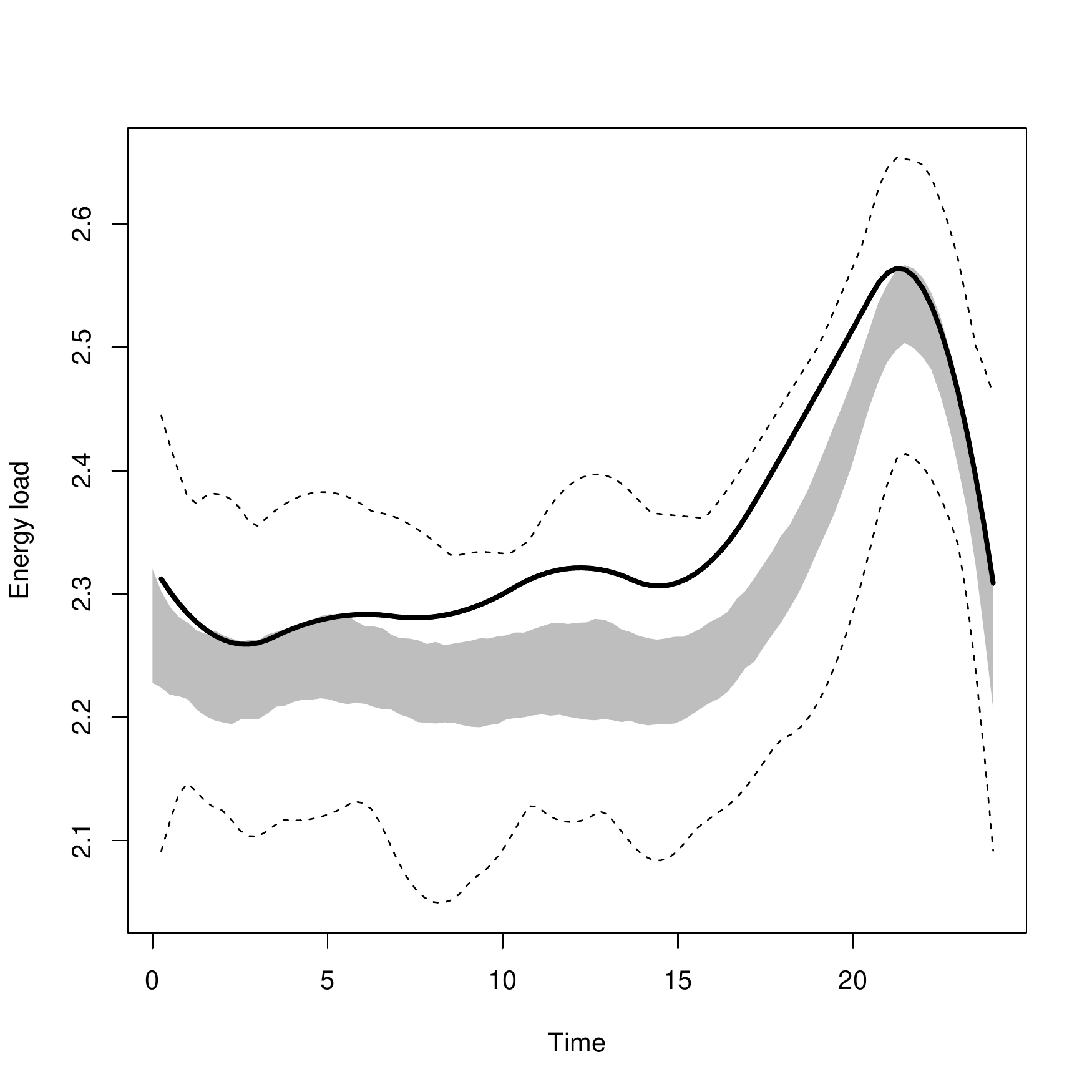}
\end{subfigure}%
\begin{subfigure}{0.5\textwidth}
\centering
  \includegraphics[width=\textwidth]{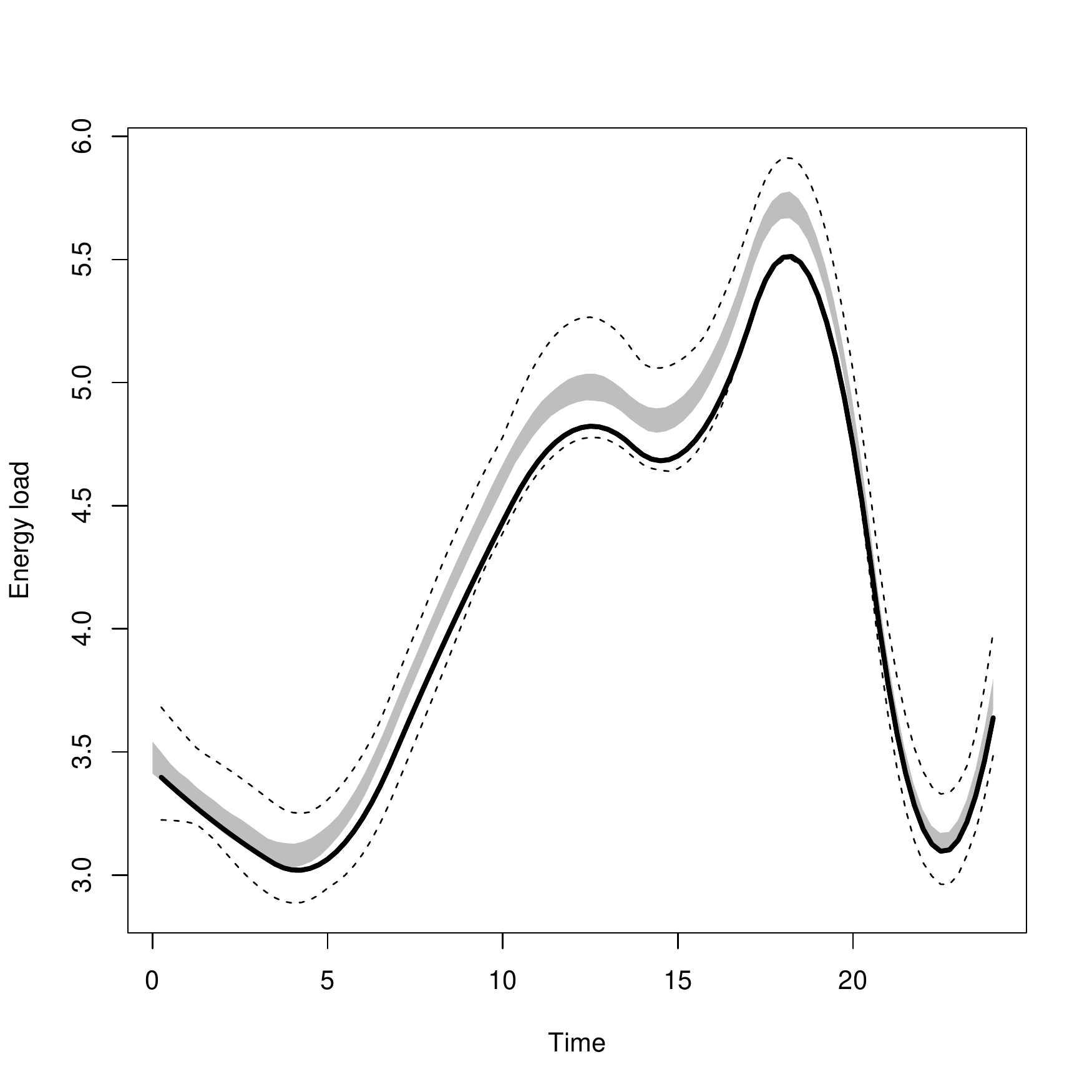}
\end{subfigure}
\begin{subfigure}{0.5\textwidth}
\centering
  \includegraphics[width=\textwidth]{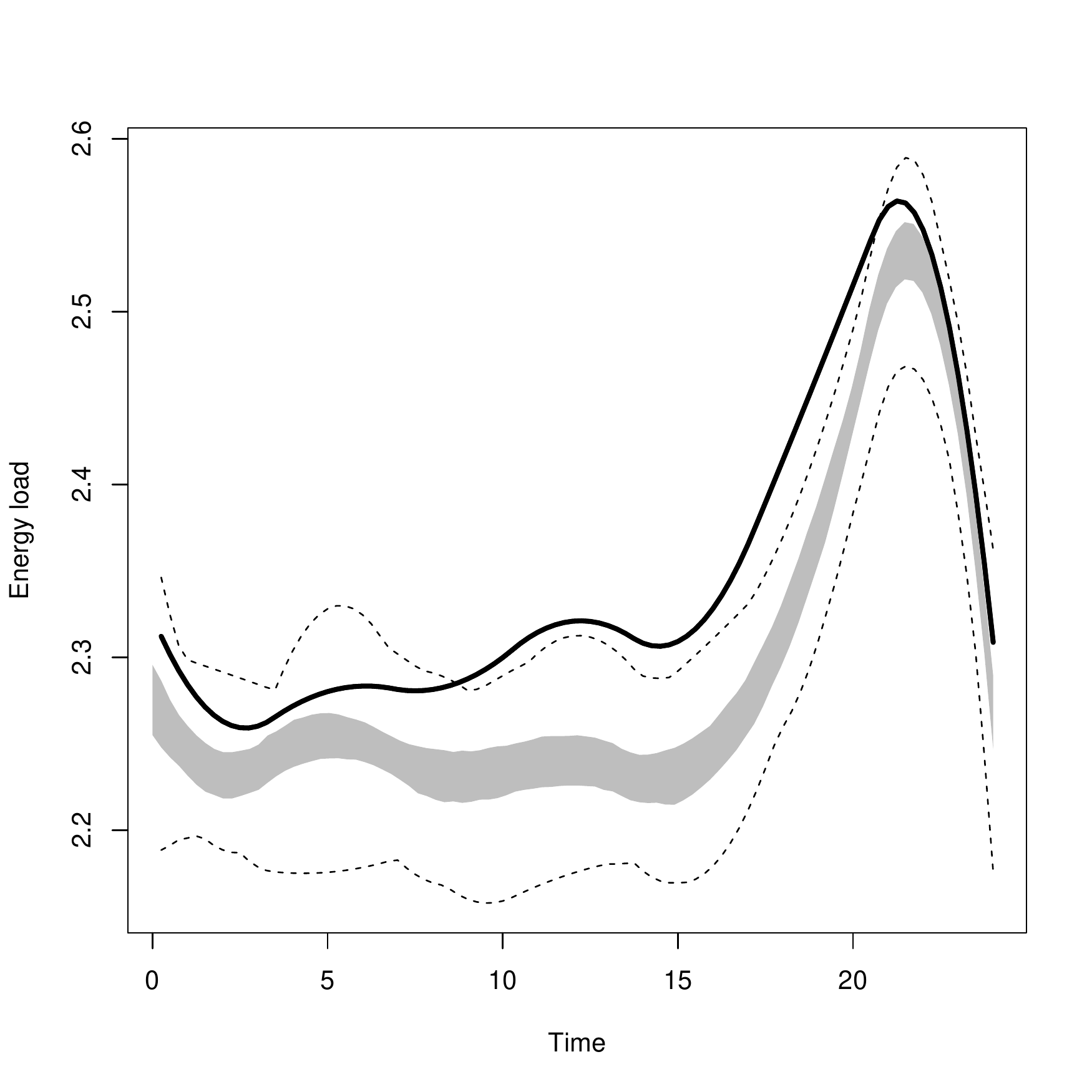}
\end{subfigure}%
\begin{subfigure}{0.5\textwidth}
\centering
  \includegraphics[width=\textwidth]{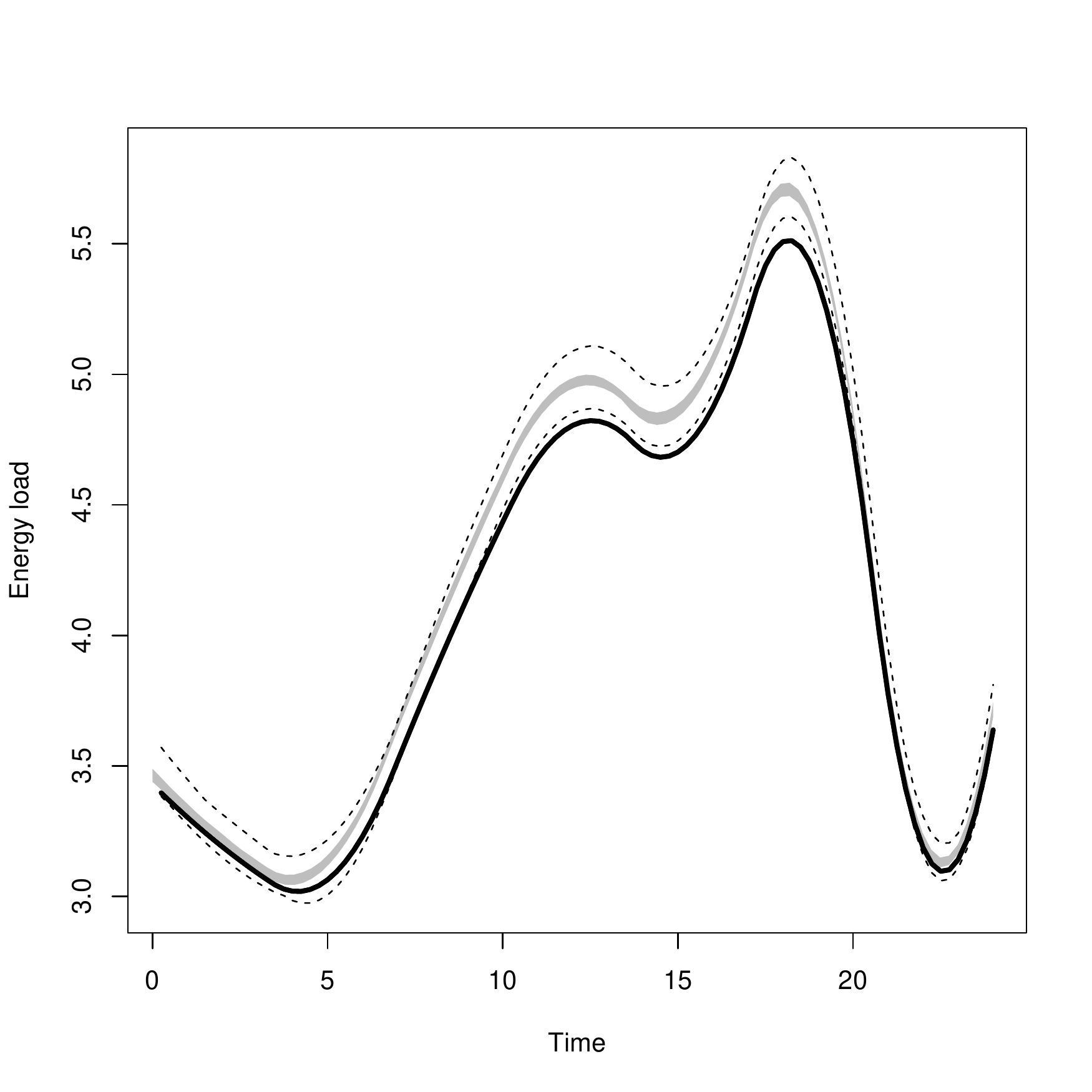}
\end{subfigure}
\caption{ Simulation study Case 3 (\textit{$\alpha_1$ is much smaller
than  $\alpha_2$ and the $M$'s are balanced}): pointwise minimum, maximum, first and third quartiles
of the 200 estimated typologies for classes $c=1$ residential (left column) and $c=2$ commercial (right column) without replicates (top row) and with replicates (bottom row) as in Figure \ref{fig:quant-case1}.}
\label{fig:quant-case3}
\end{figure}

%%%%%%%%%%%%%%%%%%%%%%%%%%%%%%%%%%%%%%%%%%%

\begin{figure}[!htb]
\centering
\begin{subfigure}{0.5\textwidth}
\centering
  \includegraphics[width=\textwidth]{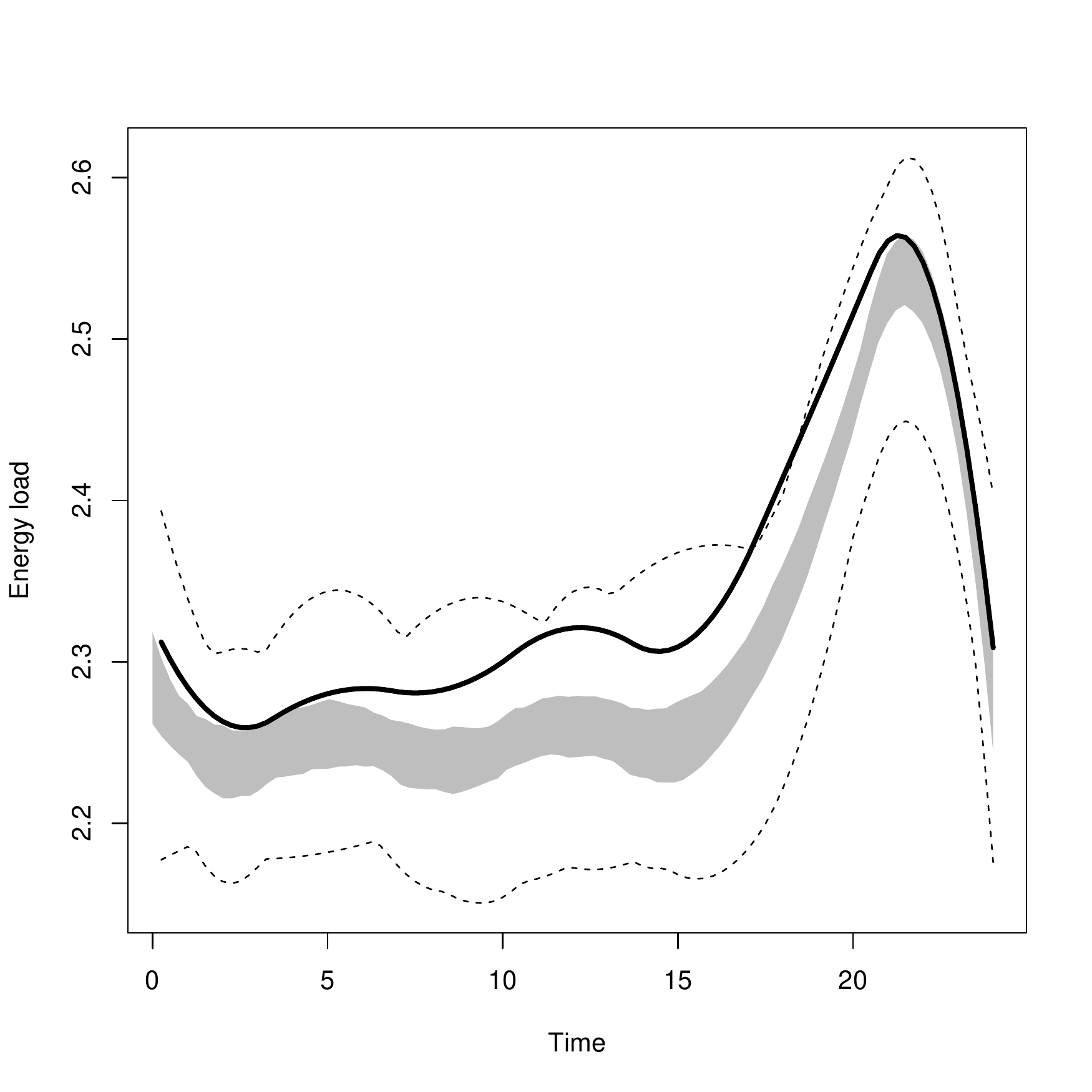}
\end{subfigure}%
\begin{subfigure}{0.5\textwidth}
\centering
  \includegraphics[width=\textwidth]{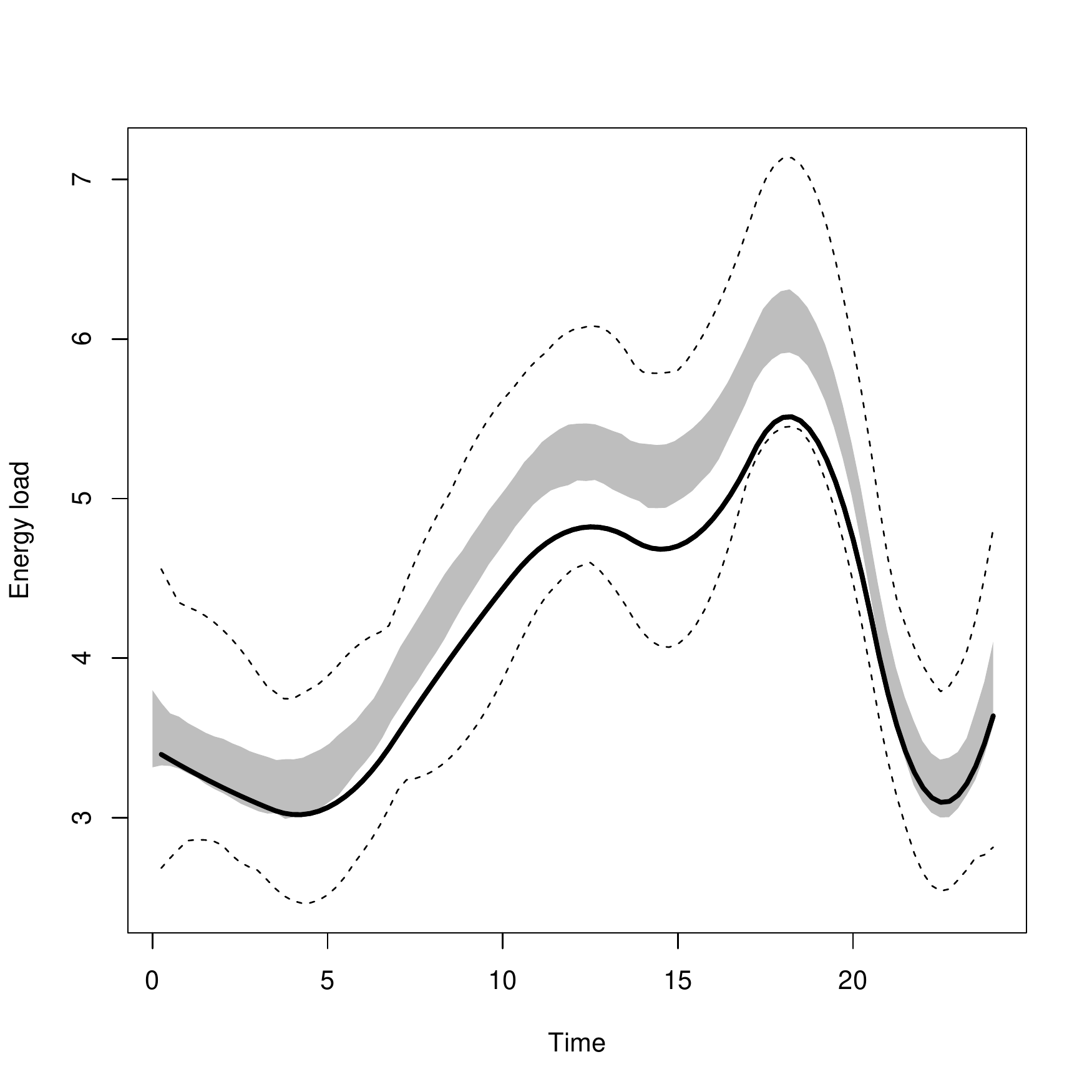}
\end{subfigure}
\centering
\begin{subfigure}{0.5\textwidth}
\centering
  \includegraphics[width=\textwidth]{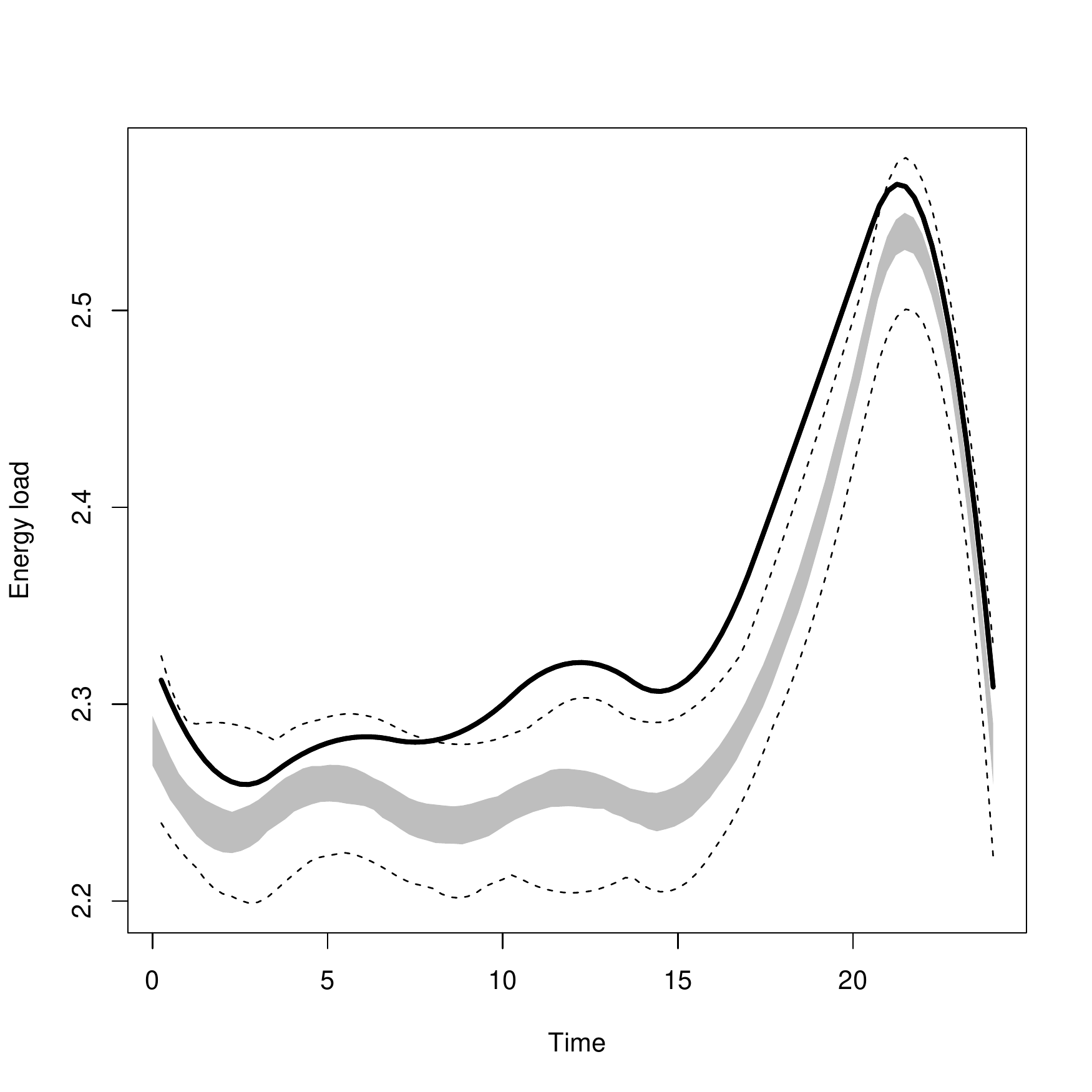}
\end{subfigure}%
\begin{subfigure}{0.5\textwidth}
\centering
  \includegraphics[width=\textwidth]{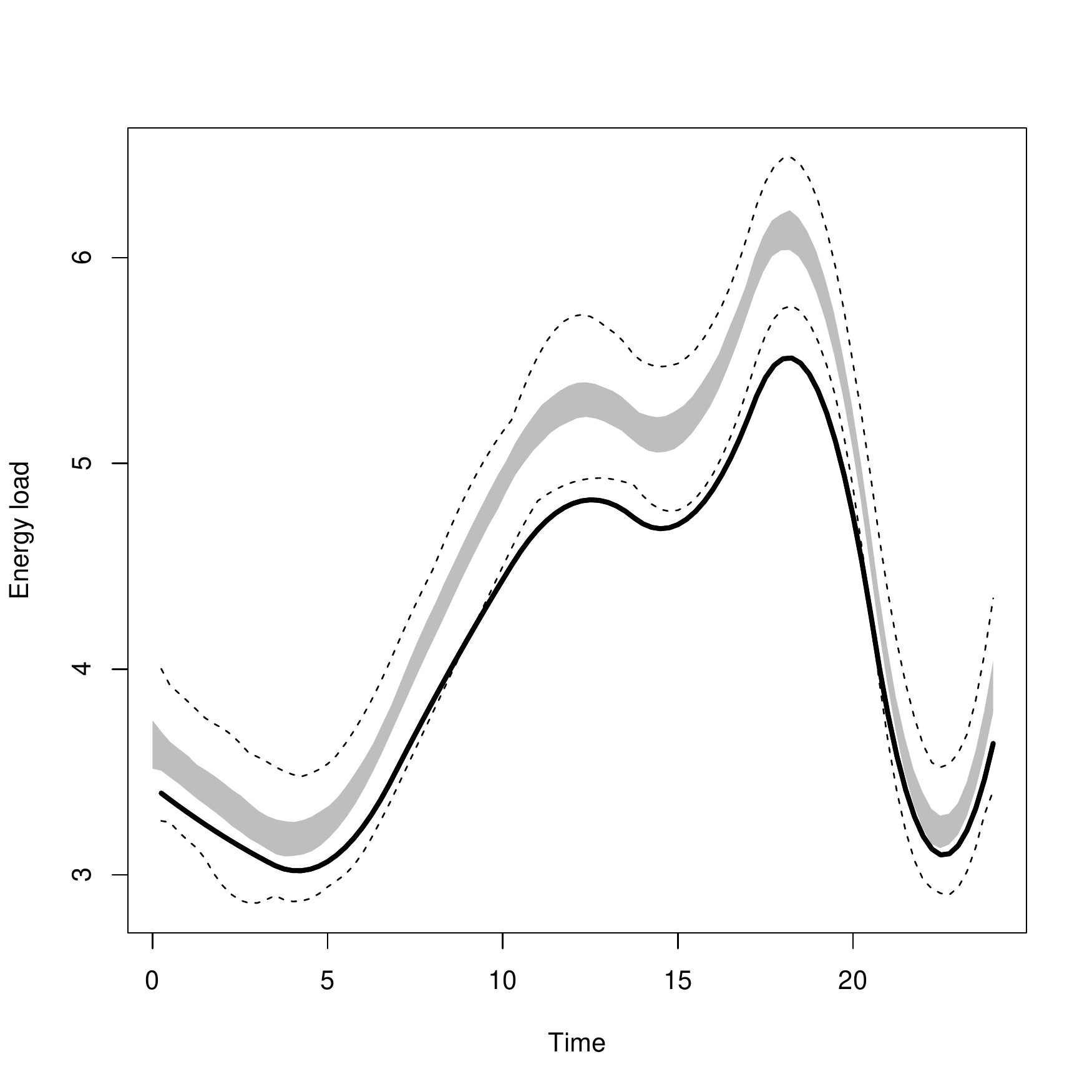}
\end{subfigure}
\caption{ Simulation study Case 4 (\textit{$\alpha_1$ is much smaller
than  $\alpha_2$ and the $M_1$'s are much bigger than the $M_2$'s}): pointwise minimum, maximum, first and third quartiles
of the 200 estimated typologies for classes $c=1$ residential (left column) and $c=2$ commercial (right column) without replicates (top row) and with replicates (bottom row) as in Figure \ref{fig:quant-case1}.}
\label{fig:quant-case4}
\end{figure}

\clearpage

%%%%%%%%%%%%%%%%%%%%%%%%%%%%%%%%%%%%%%%%%%%%%%%%%%%%%
\begin{figure}[!htb]
\centering
\begin{subfigure}{0.5\textwidth}
\centering
  \includegraphics[width=\textwidth]{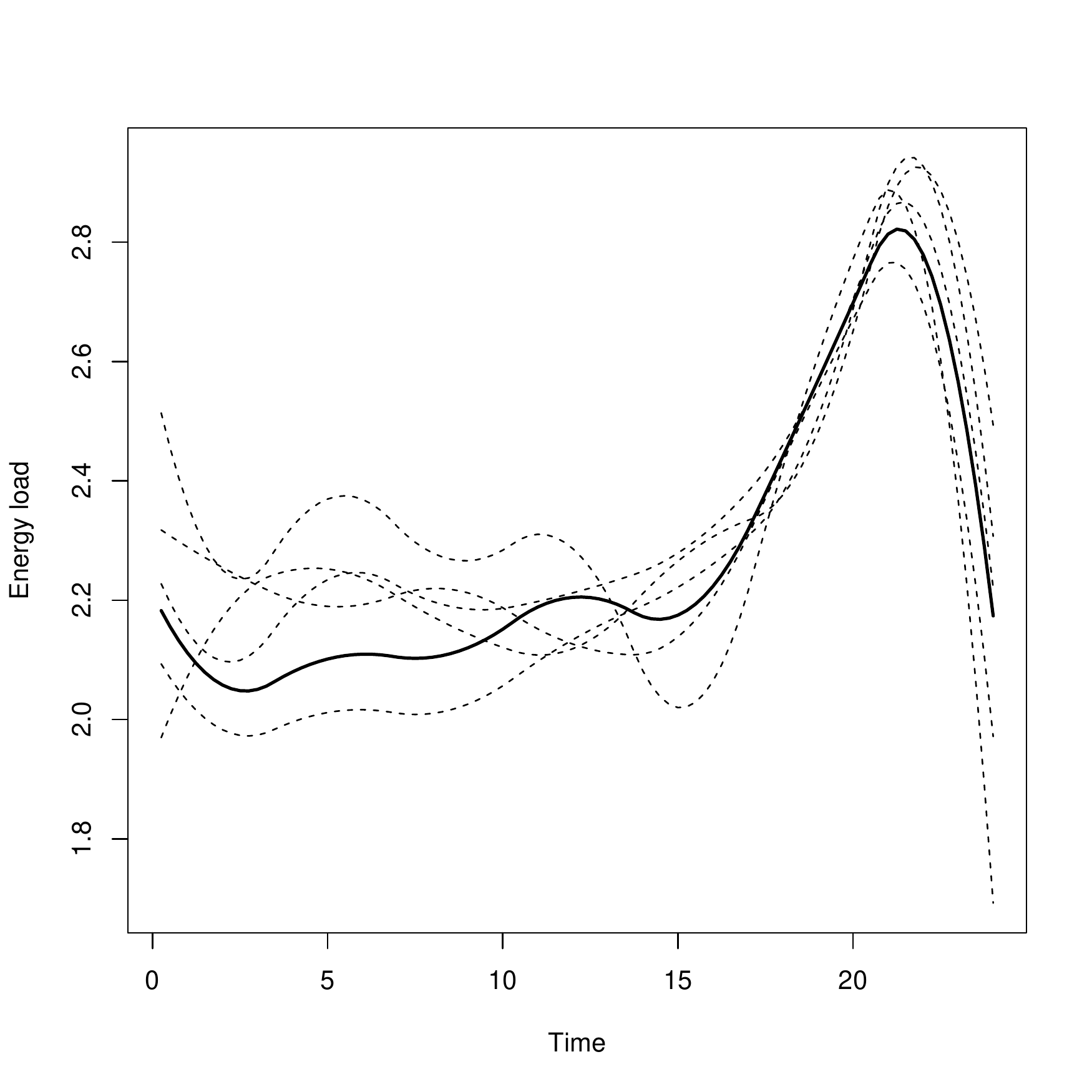}
\end{subfigure}%
\begin{subfigure}{0.5\textwidth}
\centering
  \includegraphics[width=\textwidth]{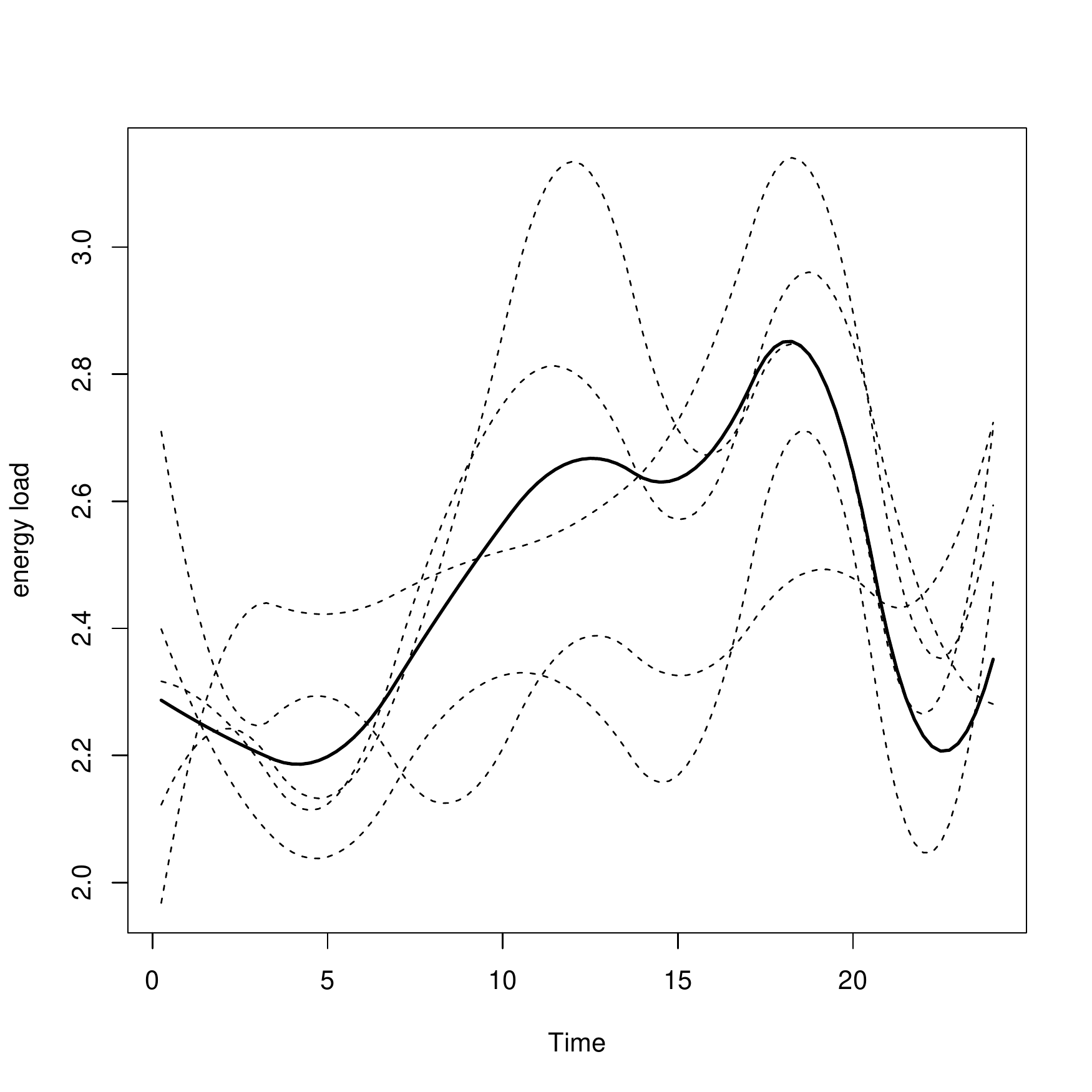}
\end{subfigure}
\caption{Simulation study Case 1 (\textit{$\alpha_1$ and $\alpha_2$ are of the same scale
and the $M$'s are balanced}): example of simulated individual level energy consumption curves (dashed curves) for 5 consumers of class 1 (left) and 5 consumers of class 2 (right) compared with the true typical curves (solid line).}
 \label{fig:simulation_data1}
\end{figure}

\clearpage

\begin{figure}
\centering
\includegraphics[width=10cm]{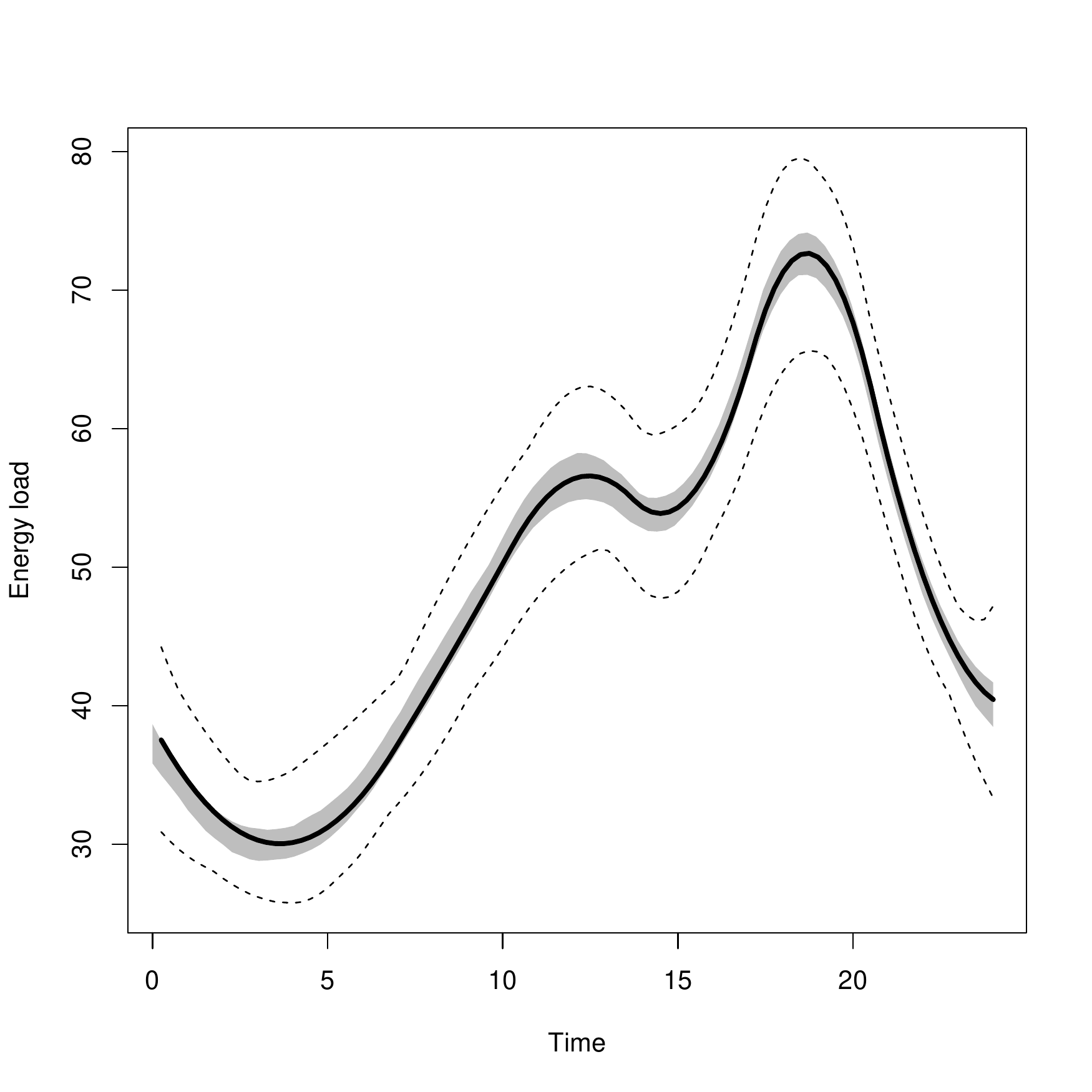}
\caption{Simulation study Case 4 (\textit{$\alpha_1$ is much smaller
than  $\alpha_2$ and the $M_1$'s are much bigger than the $M_2$'s}): estimate of the total electric load ($\hat{M}_{1i} \hat{\alpha}_1(t)
+ \hat{M}_{2i} \hat{\alpha}_2(t)$) for one of the five transformers.}
\label{fig:agregCase4_example}
\end{figure}

%%%%%%%%%%%%%%
%%% tables %%%
%%%%%%%%%%%%%%

\clearpage

\begin{table}
\centering
\renewcommand{\arraystretch}{1.2}
\begin{tabular}{ c|cc|cc|cc}
        & \multicolumn{2}{c}{Monophasic}  & \multicolumn{2}{c}{Biphasic}    &  \multicolumn{2}{c}{Commercial}
        \\
   Transformer  &   Reported & Estimated    &   Reported & Estimated &   Reported & Estimated  \\            \hline
    1           &     5      &    5         &      29    &  29       &      7     &   7     \\
    2           &     4      &    4         &      43    &  43       &      3     &   3     \\
    3           &    48      &   49         &      26    &  25       &      3     &   3     \\
\end{tabular}
\vskip 15pt
 \caption{Data set:  reported counts of consumer classes and our estimated counts for
the data.}
\label{tab:dataset}
  \end{table}

%%%%%%%%%%%%%%%%%%%%%%%%%%%%%%%%%%%%%%%%%%%%%%%%%%%%%
%%%  FOR UNDERSTANDING THE DISTRIBUTION OF THE REPORTED COUNTS

\begin{table}
\centering
\begin{tabular}{|c|ccccc|c|}
\hline
\backslashbox{True class}{Reported class}& 1 & $\cdots$  & $j$  &  $\cdots$ & $C$  &   {\rm{Row ~totals}}
  \\
  \hline
1 &$ x_{11} $ & $\cdots$  & $ x_{1j}$   & $ \cdots$    &   $x_{1C} $  & $ M_{1}$
\\
$\vdots$ &$\vdots$  &    & $\vdots$  &   & $\vdots$&  $\vdots$
\\
$c $&$x_{c1}$  & $\cdots $ &  $x_{cj} $ & $\cdots $  &  $x_{cC}  $& $ M_{c}$
\\
$\vdots$ & $\vdots$  &    & $\vdots$  &   & $\vdots$ &  $\vdots$
\\
$C $&$x_{C1} $ & $\cdots $ & $ x_{Cj}$  & $\cdots $  &  $x_{CC}$  &  $M_{C}$
\\
\hline
{\rm{Column~ totals}}  & $R_{1}$  & $ \cdots $ & $ R_{j}$  & $ \cdots$ &  $R_{C}$  & $ N $
\\ \hline
\end{tabular}
\vskip 15pt
\caption{Counts of consumer in different categories:  $x_{cj}$ is the number of consumers of type $c$ who have reported they are of class $j$.}
\label{eq:Xmat}
\end{table}

%%%%%%%%%%%%%%%%%%%%%%%%%%%%%%%%%%%%%%%%%%%%%%%%%%%%%
%%%% EXAMPLE OF CALCULATING THE H FUNCTION

\begin{table}
\centering
\begin{tabular}{c@{\hskip0.3cm}c@{\hskip 0.3cm}c@{\hskip 0.3cm}c@{\hskip 0.3cm}c@{\hskip 0.3cm}c@{\hskip 0.3cm}c@{\hskip 0.3cm}c@{\hskip 0.3cm}c@{\hskip 0.3cm}c@{\hskip 0.3cm}c@{\hskip 0.3cm}c@{\hskip 0.3cm}c@{\hskip 0.3cm}c}
%%\begin{tabular}{cccccccccccccc}
& \multicolumn{13}{c}{$m_1$}
\\
\cline{2-14}
 &
25 &
26 &
27 &
28 &
29 &
30 &
31 &
32 &
33 &
34 &
35 &
36 &
37
\\  \cline{2-14}
$\hat{H} $    & 0.000  & 0.002  & 0.007  & 0.027  & 0.075 & 0.166  & 0.256 & 0.255 & 0.143 & 0.051 & 0.014  & 0.003  & 0.000 \\
\end{tabular}
\vskip 15pt
%\end{center}
\caption{Estimates of $H(m_1, 75-m_1)$ when $R_1 = 32$ and $R_2 = 43$ for the fraud matrix in (\ref{eq:Fraud}).}
\label{table:m1.example}
\end{table}

\clearpage

%%%%%%%%%%%%%%%%%%%%%%%%%%%%%%%%%%%%%%%%%%%%%%%%%%%%%
%%%%%  FOR SIMULATION STUDIES - HOW DATA WERE GENERATED

\begin{table}
 \centering
\renewcommand{\arraystretch}{1.2}
\begin{tabular}{ c|cc|cc}
        &                  \multicolumn{2}{c}{Balanced}  &    \multicolumn{2}{c}{Unbalanced}
        \\
   Transformer               &                   Truth & Reported    &      Truth& Reported
 \\            \hline
    1     &   45   &    45         &             66   &  65
   \\
    2     &  29    &    32         &             65   & 66
   \\
   3     &  61    &    60         &             69   & 68
   \\
    4     &  24    &    28         &             62   &  63
   \\
   5     &  12    &    16         &             72   &  71
   \\ \hline
   Total  &   171   &   181    &    334  &   333
   \\
\end{tabular}
\vskip 15pt
\caption{Simulation study:  the true number of consumers of class 1 and the randomly generated reported number of consumers, as used in the simulation studies. Each transformer served a total of 75 consumers, for a total of 375 consumers.}
\label{table:true.counts}
\end{table}

  %%%%%%%%%%%%%%%%%%%%%%%%%%%%%%%%%%%%%%%%%%%%%%%%%%
  %%   RESULTS FROM SIMULATION STUDY ON ESTIMATED COUNTS

\begin{table}
\centering
\begin{tabular}{ccccccc}
%\hline
 &\multicolumn{6}{c}{${\widehat M_{1}}$} \\  \cline{2-7}
Cases &28& ${\bf{29}}$& 30& 31&$[32]$ & 33 \\ \hline
1  without replication&3&24&61&74&35&3\\
3   without replication&1&7&37&97&55&3\\
1  with 5 replications&0&0&63&135&2&0\\
3  with 5 replications&0&0&25&161&14&0\\ \hline
          \end{tabular}
   \vskip 15pt

\begin{tabular}{cccccc}
%\hline
 &\multicolumn{5}{c}{${\widehat M_{1}}$} \\  \cline{2-6}
Cases &64& ${\bf{65}}$ &  $[66]$& 67&  68  \\ \hline
2  without replication&5&63&109&21&2\\
4   without replication&0&144&56&0&0\\
2  with 5 replications&0&7&192&1&0\\
4  with 5 replications&0&170&30&0&0\\ \hline
          \end{tabular}
          \vskip 20pt

\caption{Simulation study: tables of the simulated distribution of $\hat{M}_{1}$, the estimated number of consumers of class $c=1$ (residential) in transformer
$i=$2. The top table shows the results for the cases with balanced $M$'s (Cases 1 and 3) and the bottom
table shows the results for the cases where all $M_{1}$'s are much larger than the $M_{2}$'s (Cases 2 and 4) . In each table,
the column heading  contains different possible estimate values with the true value of $M_1$ in bold font and the reported value within brackets.
Each of the remaining rows corresponds to a different simulation scenario.  Each row contains  the number of  estimates of $M_1$ (out of the 200 estimates) that are equal to the associated column heading.
}
\label{tab:traf2-bal}
\end{table}

\clearpage

\end{document}